\documentclass[12pt]{IEEEtran}
\usepackage{amsmath,amssymb,verbatim}
\newcommand{\mysectionapp}[1]{\section{Appendix \Alph{section}: #1}}

\newtheorem{theorem}{Theorem}%[section]
\newtheorem{lemma}{Lemma}%[section]
\newtheorem{definition}{Definition}%[section]
\newtheorem{corollary}{Corollary}%[section]
\newtheorem{remark}{Remark}%[section]
\newtheorem{example}{Example}%[section]

\newcommand{\enproof}{\hfill $\Box$ \vspace*{1ex}}
\newcommand{\enlem}{\hfill $\Diamond$ \vspace*{1ex}\end{lemma}}
\newcommand{\closedef}{\hfill $\Diamond$ \end{definition}}
\newcommand{\enth}{\hfill $\Diamond$ \end{theorem}}
\newcommand{\encor}{\hfill $\Diamond$ \end{corollary}}
\newcommand{\enprop}{\hfill $\Diamond$ \end{proposition}}
\newcommand{\encond}{\hfill $\Diamond$ \end{condition}}
\newcommand{\exam}[1]{\begin{example}\label{ex:#1}}
\newcommand{\enexam}{\QED\end{example}}
\newcommand{\beremark}[1]{\begin{remark}\label{rmk:#1}}
\newcommand{\enremark}{\enproof}

\newcommand{\mymathbb}[1]{{\mathbb #1}} 
\newcommand{\mymathsf}[1]{{\mathsf #1}} %IEICE, IEEE
\newcommand{\cA}{{\cal A}}
\newcommand{\sA}{\mymathsf{A}}

\newcommand{\sB}{\mymathsf{B}}

\newcommand{\cC}{{\cal C}}
\newcommand{\bC}{\mymathbb{C}}

\newcommand{\sF}{\mymathsf{F}}
\newcommand{\myF}{{\sF}}

\newcommand{\rsC}{\mymathsf{S}}
\newcommand{\rsCgen}{\mymathsf{T}}
\newcommand{\sH}{\mymathsf{H}}
\newcommand{\cI}{{\cal I}}

\newcommand{\sL}{\mymathsf{L}}
\newcommand{\cM}{{\cal M}}

\newcommand{\sN}{\mymathsf{N}}
\newcommand{\cP}{{\cal P}}
\newcommand{\sP}{\mymathsf{P}}

\newcommand{\cR}{{\cal R}}

\newcommand{\cT}{{\cal T}}
\newcommand{\cU}{{\cal U}}

\newcommand{\cX}{{\cal X}}
\newcommand{\cY}{{\cal Y}}
\newcommand{\cW}{{\cal W}}
\newcommand{\cZ}{{\cal Z}}
\newcommand{\bZ}{\mymathbb{Z}}
\newcommand{\vNe}{S} %{H}

\renewcommand{\phi}{\varphi}
\renewcommand{\subset}{\subseteq}
\renewcommand{\tilde}{\widetilde}
\renewcommand{\hat}{\widehat}
\renewcommand{\Bar}{\overline}

\newcommand{\dss}{\displaystyle}
\newcommand{\mbm}[1]{\mbox{\boldmath $#1$}}

\newcommand{\defeq}{\stackrel{\rm def}{=}}
\newcommand{\mateq}{\stackrel{\rm m}{\sim}}

\newcommand{\SINT}{\mymathbb{Z}}

\newcommand{\field}{\mymathbb{F}}
\newcommand{\Capa}{\mymathsf{C}}

\newcommand{\Prob}{{\rm Pr}}

\newcommand{\transp}{^{\rm T}}

\newcommand{\tnsr}{\otimes}

\newcommand{\lag}{\langle}
\newcommand{\rag}{\rangle}
\newcommand{\crd}[1]{|#1|}
\newcommand{\bra}[1]{\lag #1 |}
\newcommand{\ket}[1]{| #1 \rag}
\newcommand{\indc}{{\bf 1}}
\newcommand{\syp}[2]{( #1,  #2 )_{\rm sp}}

\newcommand{\dmn}{d}
\newcommand{\Hch}{{\sH}}
\newcommand{\Hgn}{{\sH}}
\newcommand{\Bop}{\sL}
\newcommand{\Hcd}{\cC}

\newcommand{\Fav}{F_{\rm a}}

\newcommand{\imu}{\sqrt{-1}} %{{\rm i}}
\newcommand{\Ebasis}{\sN}
\newcommand{\Ebe}{N}
\newcommand{\Ebeh}[1]{N^{(#1)}}
\newcommand{\Ebeg}[1]{N^{[#1]}}
\newcommand{\Ebesp}[1]{N^{\langle #1 \rangle}}
\newcommand{\ketbe}[1]{\ket{#1}}
\newcommand{\phasebe}{\omega}
\newcommand{\Ebasisch}{\sN} %{\tilde{\sN}}
\newcommand{\Xbe}{X}
\newcommand{\Zbe}{Z}

\newcommand{\Enk}[2]{E_{#1,#2}}
\newcommand{\Icr}{J} 
\newcommand{\Fbar}{\overline{F}}
\newcommand{\Aso}{\sA}  %for self-orthogonal
\newcommand{\Acn}{\sA}  %for conditional
\newcommand{\Bcn}{\sB}  %for conditional
\newcommand{\CPex}{\cM}
\newcommand{\CPexO}{M}
\newcommand{\Cso}{L}
\newcommand{\cnc}[2]{\mymathsf{cat}({#1},{#2})}

\newcommand{\varu}{u} %used to be v
\newcommand{\varupr}{\tilde{u}} %used to be u''
\newcommand{\varv}{v} %used to be w
\newcommand{\varz}{\sigma}

\newcommand{\varb}{b} %r
\newcommand{\vard}{b'}
\newcommand{\tvarz}{w} %4switch
\newcommand{\tvarw}{z} %4switch
\newcommand{\varl}{m} %used to be l
\newcommand{\varnin}{n} %4 inner codes %4switch used to be \nu
\newcommand{\varkin}{k} %4 inner codes %4switch used to be \kappa
\newcommand{\varN}{\nu}   %used be n 
\newcommand{\varK}{\kappa}  %used be k 
\newcommand{\varm}{m} %used be a'
\newcommand{\zpr}{\tilde{z}}
\newcommand{\varzpr}{\sigma}
\newcommand{\baseChoi}[1]{r_{#1}} %used be b_

\newcommand{\rvX}{\mymathsf{X}}
\newcommand{\rvY}{\mymathsf{Y}}

\newcommand{\spn}{\mymathsf{span}\,}
\newcommand{\Gin}{G}
\newcommand{\spGin}{\spn \Gin}
\newcommand{\yg}{Y_{G}}
\newcommand{\Cin}{\Cso}
\newcommand{\Cout}{\Cso_{\rm out}}
\newcommand{\Pin}{P_{\Cso}}
\newcommand{\Pinpr}{P_{\Cso'}}
\newcommand{\cset}[2]{\Gamma{(#2)}}
\newcommand{\Lcset}[2]{\Lambda{(#2)}}
\newcommand{\Proj}{\Sigma} %{\Lambda}

\newcommand{\coch}{\gamma} %{\alpha}
\newcommand{\cochith}[2]{\coch_{#2}(#1)}
\newcommand{\ghb}[1]{\Bar{#1}}
\newcommand{\smlsum}{\Sigma}
\newcommand{\Id}{{\rm I}}
\newcommand{\pdsm}{\times} %{*} %{\cdot}
\newcommand{\tcQ}{\cP} %used be \cQ
\newcommand{\cndV}{\cW} %\cV
\newcommand{\cardcndV}{W_{\varN}}
\newcommand{\cardVsh}[2]{T^{\nu}_{#2}(#1)}
\newcommand{\varW}{V} %used to be W
\newcommand{\rvZ}{\mbm{z}}
\newcommand{\rvV}{\mbm{v}}
\newcommand{\varsp}{s} %used to be t
\newcommand{\varrc}{x} %used to be r
\newcommand{\vartpr}{s} %used to be s and then t'
\newcommand{\varM}{m} 
\newcommand{\varDK}{K} 
\newcommand{\varNK}{M} 

\title{%
Information Rates Achievable with Algebraic Codes on Quantum Discrete Memoryless Channels}

\author{Mitsuru Hamada\\[1ex]
{\normalsize Quantum Computation and Information Project (ERATO)\\
     Japan Science and Technology Corporation\\
              201 Daini Hongo White Bldg.,\
      5-28-3, Hongo, Bunkyo-ku, Tokyo 113-0033, Japan\\
E-mail: {\tt mitsuru@ieee.org}}
}
\begin{document}

\maketitle

\begin{abstract}
The highest information rate at which 
quantum error-correction schemes work reliably on a channel,
which is called the quantum capacity, is proven
to be lower bounded by the limit of the quantity
termed coherent information maximized over
the set of input density operators which are proportional to
the projections onto the code spaces of symplectic stabilizer codes.
Quantum channels to be considered
are those subject to independent errors
and modeled as tensor products of copies of a 
completely positive linear map on a Hilbert space of finite dimension,
and the codes that are proven to have the desired performance  %ver2
are symplectic stabilizer codes.
On the depolarizing channel, this work's bound is actually
the highest possible rate at which symplectic stabilizer codes work reliably.
\end{abstract}


\section{Introduction \label{ss:intro}}

The problem of determining
the capacity of quantum channels
was posed by Shor~\cite{shor95} in the
first paper on quantum error-correcting codes (quantum codes, or
codes, hereafter).
He discussed it in the context of
preservation of quantum states, which are to be used 
for quantum computation in the presence of quantum noise.
There is a known upper bound on the quantum capacity
based on the quantity called coherent information,
and some authors conjecture that this bound is tight~\cite{barnum98}, \cite[Section~VI]{barnum00}, \cite{HolevoWerner01,holevo01s,HPreskill01}.
On the other hand, known lower bounds appear to have left
much room for improvement.
For example, on the capacity of 
the depolarizing channel, which suffers uniform depolarization
and can be specified by Kraus operators
$\sqrt{1-p} I, \sqrt{p/3} X,\sqrt{p/3} Y,\sqrt{p/3} Z$
with $I$ and $X, Y, Z$ being the identity and Pauli operators, respectively,
the highest lower bound known is $1-h(p)-p\log_2 3$ for a wide range of $p$,
where $h$ is the binary entropy function~\cite{bennett96m,gottesmanPhD,preskillLNbook0,hamada01e,hamada01g}.
Shor and Smolin~\cite{ss97} argued this bound is not tight
showing the existence of concatenated quantum codes that
slightly go beyond it for a limited range of $p$,
which revealed a remarkable feature of the issue of the quantum capacity.
While their work and the subsequent analysis of
DiVincenzo and these authors~\cite{dss98} abounded with suggestions,
their code construction was apparently restricted, and
explorations into the general nature behind their code construction
and further analysis were awaited~\cite{dss98}.

The aim of this work is to give a more general lower bound
which partially closes the gap between the upper and lower bounds,
at least qualitatively.
The bound to be presented is expressed as 
the limit of coherent information maximized over
the set of input density operators which are proportional to
the projections onto the code spaces of standard algebraic quantum codes.
This limit closely resembles the known upper bound on the capacity,
which is the one defined in the same way but with the restriction on 
input density operators removed.
The result is obtained by developing Shor and Smolin's idea
on the basis of the geometric property of quantum codes,
and incorporating a methodology from classical information theory.
In other words, this work establishes an exponential lower bound on %ver2
the highest fidelity of concatenated quantum codes
used on a memoryless channel in an elementary enumerative manner
employing the method of types~\cite{csiszar_koerner,csiszar98}.
%and also using algebraic properties of quantum codes.
This fidelity bound then
%improves that of the previous works of this author~\cite{hamada01e,hamada01g},
%and
gives the new lower bound on the quantum capacity of memoryless channels.
Unlike Shor and Smolin's or DiVincenzo and these authors'
coding schemes~\cite{ss97,dss98}, the codes in this work
do not rely on purification protocols and
fall in the class of standard `in-place' quantum ones called stabilizer
codes, which would be desirable
for its simplicity of coding processes %~\cite{bennett96m}
and %possible implementability for use
the possibility to be used
in quantum computation%
~\cite{shor95f,steane99,gottesmanPhD}.
%nielsen_chuang}.
Moreover, for the depolarizing channel, which
has often been adopted as a channel model for analysis of quantum codes,
it will be shown that this bound on the capacity is the highest
possible that can be attained with standard quantum codes.

Concatenated quantum codes are, 
in a sense, analogous to classical concatenated codes~\cite{forney} and
form a subclass of the class of standard algebraic quantum codes,
which are called stabilizer, additive or symplectic codes
in the literature~\cite{gottesman96,gottesmanPhD,crss97,crss98,rains99,hamada01e}.
While the term stabilizer codes is prevailing,
it would rather be called {\em symplectic (quantum) codes}\/ %~\cite{rains99}
or {\em symplectic stabilizer codes}\/
with emphasis on the role of symplectic geometry in this work.
%in designing these codes.
%and the former term might deserve applications to wider 
A symplectic quantum code is a simultaneous eigenspace
of a set of commuting operators, which is called a stabilizer.
A stabilizer is obtained by constructing
a code over a finite field which is self-orthogonal
with respect to a symplectic bilinear form and then transforming
it into operators on a Hilbert space through a one-to-one correspondence
(a projective representation).
A stabilizer of a concatenated quantum code, 
which will simply be called a concatenated code in what follows,
is obtained by concatenating two such self-orthogonal codes
and putting it through the representation.
We refer to these two codes, or the corresponding quantum codes,
an inner code and an outer one following Forney's terminology~\cite{forney}.
Shor and Smolin's concatenated code uses an inner code with restricted
parameters. Namely, their inner code is an $[[\varnin,\varkin=1]]$ code,
where 
an $[[\varnin,\varkin]]$ code is a $2^{\varkin}$-dimensional
subspace of the tensor
product of $n$ copies of a two-dimensional Hilbert space.
%(in the case of $\varkin=1$ below).
This paper develops DiVincenzo, Shor and Smolin's analysis~\cite{ss97,dss98}
to include that of
concatenated codes with general inner $[[\varnin,\varkin]]$ codes
with $1\le\varkin\le\varnin$.
%by which the highest information rate is achievable.

To the still ongoing development of the theory of
quantum channel coding, %capacity issues, 
there have been many authors' contributions.
Good surveys on these problems have been given in
\cite{barnum00} and \cite{holevo01s}.
An incomplete list of contributions after either of these surveys includes
\cite{GottesmanKP01,HPreskill01,HolevoWerner01,ABargKnillL00,FNagaoka98,king01a,king02,AHolevo01,shor02a,bs01ea,holevo01ea,KeylWerner02}.
%holevo00,BurnashevHolevo97}. 
Especially, we have witnessed
the determination of the entanglement-assisted capacity~\cite{bs01ea} (see also \cite{holevo01ea}) and the settlement of the additivity problem of
the classical capacity for several classes of channels~\cite{king01a,king02,shor02a}
while these are not the capacity which this paper will be concerned with.
Nor will it discuss continuous channel 
models such as quantum Gaussian channels~\cite{GottesmanKP01,HPreskill01,HolevoWerner01}.

The argument below proceeds as follows.
After stating the result (Section~\ref{ss:cap}), 
we first recapitulate the framework of symplectic codes
in a self-contained manner
assuming no formidable prerequisite such as knowledges on %acquaintance with
representation theory, though a few basic facts
from geometric algebra~\cite{artin,grove} are used (Section~\ref{ss:sc}).
Then, concatenated codes are explicated in this framework
(Section~\ref{ss:cat}), and
the lower bound on the capacity is proven in an elementary manner
with the aid of the method of types (Section~\ref{ss:proof} and Appendix~\ref{ss:proof4Lemma5}).
Thereafter, 
it is shown that this bound is the highest 
possible on the depolarizing channel 
if we restrict the coding schemes to symplectic stabilizer
codes (Section~\ref{ss:cond_cap}).
Finally, some remarks are given on the case of general channels and so on
(Sections~\ref{ss:gen} and \ref{ss:conc}).
Appendices are given to prove lemmas and a theorem.

%\section{Main Result \label{ss:cap}}
\section{Quantum Capacity and New Lower Bound \label{ss:cap}}

As usual, all quantum channels
and decoding (state-recovery) operations in coding systems
are described in terms of 
{\em trace-preserving completely positive}\/ (TPCP) linear maps
~\cite{kraus71,choi75,schumacher96,barnum00,nielsen_chuang}.
Given a Hilbert space $\Hgn$ of finite dimension,
let $\Bop(\Hgn)$ denote the set of linear operators on $\Hgn$. 
In general, every completely positive (CP)
linear map $\CPex: \Bop(\Hgn) \to \Bop(\Hgn)$ 
has an operator-sum representation
$\CPex(\rho) = \sum_{i\in\cI} \CPexO_i \rho \CPexO_i^{\dagger}$ with some
$\CPexO_i\in\Bop(\Hgn)$, $i\in\cI$~\cite{kraus71,choi75,schumacher96,nielsen_chuang}.
When $\CPex$ is specified by a set of operators
$\{ \CPexO_i \}_{i\in\cI}$ %, which is not unique, 
in this way, we write
$\CPex \sim \{ \CPexO_i \}_{i\in\cI}$.%
\footnote{Here is a word about notations on ordered sets. 
Though sets of the form $\{ x_i \}_{i\in\cI}$
represent ordered ones with arbitrarily fixed orderings in principle, %ver3
the set operation $\cup$ will sometimes be applied to these
as in $\{ x_i \}_{i\in\cI}\cup\{ y_i \}_{i\in\cI'}$
if the order %of elements 
really does not matter as in operator-sum representations.}

Hereafter, $\Hch$ denotes an arbitrarily fixed Hilbert space
of dimension $\dmn$, which is a prime number.
A quantum memoryless channel is a
TPCP linear map $\cA  : \Bop(\Hch) \to \Bop(\Hch)$.
A memoryless channel $\cA$ is supposed to act
on a state or a density operator $\rho$ in $\Bop(\Hch^{\tnsr n})$
as $\cA^{\tnsr n}(\rho)$.
A pair $(\Hcd_n, \cR_n)$ consisting of a subspace
$\Hcd_n \subset \Hch^{\tnsr n}$
and a TPCP linear map
$ 
\cR_n: \Bop(\Hch^{\tnsr n}) \to \Bop(\Hch^{\tnsr n})
$,
which is supposed to serve as a recovery operator,
is called a {\em (quantum) code},\/ %for $\Hch^{\tnsr n}$,
its {\em information rate}\/ (or simply rate) 
is defined to be $n^{-1} \log_{\dmn} \dim \Hcd_n$,
and its performance is evaluated in terms of {\em minimum fidelity}\/%
~\cite{KnillLaflamme97,dss98,barnum00}
\begin{equation}\label{eq:defF}
F(\Hcd_n, \cR_n\cA^{\tnsr n}) = \min_{ \ket{\psi} \in \Hcd_n } 
\lag\psi| \cR_n\cA^{\tnsr n}(|\psi\rag \lag\psi|) |\psi\rag,
\end{equation}
where $\cR_n\cA^{\tnsr n}$ denotes the composition of 
$\cA^{\tnsr n}$ and $\cR_n$.
Throughout, bras $\bra{\cdot}$ and kets $\ket{\cdot}$ are
assumed normalized.
A subspace $\Hcd_n$ alone is also called a code
assuming implicitly some recovery operator.

%We confine ourselves to a special class of channels.
For simplicity, 
we will work on a special class of channels that are specified as follows
though the lower bound to be presented is applicable to general channels
(Section~\ref{ss:gen}).
Fix an orthonormal basis 
$\{ \ketbe{0},\dots, \ketbe{\dmn-1} \}$ of $\Hch$.
Put $\cX=\{0,\dots,\dmn-1\}^2$
and 
\begin{equation}\label{eq:error_basis0}
\Ebe_{(i,j)} = \Xbe^i \Zbe^j, \quad (i,j)\in \cX,
\end{equation}
where 
$\Xbe, \Zbe \in \Bop(\Hch)$ are Weyl's unitary operators defined by
\begin{equation}\label{eq:error_basis}
\Xbe \ketbe{j}  = \ketbe{(j-1) \bmod \dmn\, }, \quad
\Zbe \ketbe{j} = \phasebe^ j \ketbe{j}
\end{equation}
with $\phasebe$ being a primitive $\dmn$-th root of unity%
~\cite{weyl31,schwinger60,knill96a,knill96b}.
Observe the relation
\begin{equation}\label{eq:XZomega}
\Xbe\Zbe = \phasebe \Zbe\Xbe.
\end{equation}
The set $\Ebasis =\{ \Ebe_{u} \}_{u\in\cX}$ is a basis of $\Bop(\Hch)$
and could be viewed as
a generalization of the Pauli operators (including the identity).
We will treat channels that can be written as
$\cA \sim \{ \sqrt{P(u)} \Ebe_{u} \}_{u\in\cX}$,
which will be called Pauli channels or $\Ebasis$-channels,
where $P$ is a probability distribution on $\cX$.
From the basis $\{ \Ebe_{(i,j)} \}$ of $\Bop(\Hch)$,
we obtain a basis $\Ebasis_n = \{ \Ebe_x \}_{x \in \cX^n}$
of $\Bop(\Hch^{\tnsr n})$,
where
$
\Ebe_x = \Ebe_{x_1} \tnsr \dots \tnsr \Ebe_{x_n}
$
for $x=(x_1,\dots,x_n)\in \cX^n$.
With this notation, 
we can write
$\cA^{\tnsr n}(\rho)=\sum_{x\in\cX^n} P^n(x) \Ebe_x \rho \Ebe_x^{\dagger}$
for $\cA \sim \{ \sqrt{P(u)} \Ebe_{u} \}_{u\in\cX}$,
where for a probability distribution $Q$ on a finite set $\cY$,
the product measure $Q^n$ is defined by $Q^n(y)=\prod_{i=1}^{n} Q(y_i)$,
$y=(y_1,\dots,y_n)\in\cY^n$.
Since %the resulting state
$\cA^{\tnsr n}(\rho)=\sum_{x\in\cX^n} P^n(x) \Ebe_x \rho \Ebe_x^{\dagger}$
can be viewed as the probabilistic mixture of states
$\Ebe_x \rho \Ebe_x^{\dagger}$ with probabilities $P^n(x)$,
we often say that an error $\Ebe_x$ occurs with probability $P^n(x)$
describing the action of the Pauli channel $\cA$.

We can define the capacity of a quantum channel
as in classical information theory.

\begin{definition} \label{def:cap}
Let $F_{n,k}^{\star}(\cA^{\tnsr n} )$ denote the
supremum of
$F(\Hcd, \cR\cA^{\tnsr n})$ such that there exists a code 
$(\Hcd\subset\Hch^{\tnsr n}, \cR)$
with $\log_{\dmn} \dim \Hcd \ge k$, where $n>0$ is an integer
and $k$, $0 \le k \le n$, is a real number.
The supremum of nonnegative numbers $R$ satisfying
\[
\limsup_{n\to\infty} F_{n,Rn}^{\star}(\cA^{\tnsr n})=1
\]
is called the {\em quantum capacity}\/ of $\cA$ and
denoted by $\Capa(\cA)$.
\closedef

{\em Remarks.}\/ 
The term quantum capacity is used when one needs to distinguish it from
other capacities (such as classical capacity) 
of a quantum channel~\cite{BennettShor98}.
Variations exist in definitions of capacity concepts.
Especially, besides the definition of quantum capacity above,
there exists a seemingly different one
based on entanglement fidelity, but actually they are the same~\cite{barnum00}.
In the above definition, the stipulation
$\limsup_{n} F_{n,Rn}^{\star}(\cA^{\tnsr n})=1$ 
may be replaced by
$\liminf_{n} F_{n,Rn}^{\star}(\cA^{\tnsr n})=1$;
by slightly modifying
the proof of Theorem~\ref{th:main} to be presented below,
it easily follows
that the main result holds true if the limit superior 
is replaced by the limit inferior. 
\enremark

Given a probability distribution $Q$ on $\cY\times\cZ$, we define
$\Bar{Q}$ by
\[
\Bar{Q}(u) = \sum_{v \in \cZ}Q(u,v),\quad u \in \cY,
\]
which is a marginal distribution of $Q$, and
define $\overleftarrow{Q}(\cdot | u)$ 
by 
\[
\overleftarrow{Q}(v|u)=Q(u,v)/\Bar{Q}(u),\quad v\in\cZ.
\]
for $u\in\cY$ with $\Bar{Q}(u)>0$
while $\overleftarrow{Q}(\cdot|u)$ is {\em undefined}\/ for $\Bar{Q}(u)=0$.
The classical Kullback-Leibler information
(informational divergence or relative entropy)
is denoted by $D$ and (conditional) Shannon entropy by $H$.
%~\cite{csiszar_koerner,csiszar98,ccc}.
Specifically, for probability distributions $P$ and $Q$ on a finite set $\cY$,
we define $D(P||Q)$ by $D(P||Q)=\sum_{u\in\cY} P(u) \log_{\dmn} [P(u)/Q(u)]$
and $H(Q)$ by $H(Q)= - \sum_{u\in\cY} Q(u) \log_{\dmn} Q(u)$.
In addition, for a stochastic matrix $P(\cdot|\cdot)$, %ver3
%\footnote{The matrix $(p_{uv})$ corresponding to $P(v|u)$, $u,v\in\cX$,
%is specified by $p_{uv}=P(v|u)$~\cite{csiszar_koerner}.}
i.e., a set of probability distributions $P(\cdot|u)$, $u\in\cY$,
and a probability distribution $p$ on $\cY$, 
we define $H(P|p)$ by
\[
H(P|p)= - \sum_{u\in\cY:\, p(u)>0}\sum_{v} p(u)P(v|u) \log_{\dmn} P(v|u),
\]
which is called the entropy of $P(\cdot|\cdot)$ conditional on $p$.
By convention, we assume
$\log(a/0) = \infty$ for $a>0$ and $0\log 0= 0 \log (0/0) = 0$.

For a density operator $\rho \in \Bop(\Hch')$ and 
a TPCP map $\cA': \Bop(\Hch') \to \Bop(\Hch')$,
the coherent information $I_{\rm c}(\rho, \cA')$ is defined by
\[
I_{\rm c}(\rho, \cA')=
\vNe\big(\cA'(\rho)\big) - \vNe\big( [\Id \tnsr \cA'](\ket{\Psi} \bra{\Psi})\big),
\]
where $\vNe(\sigma)$ denotes the von Neumann entropy of $\sigma$,
$\Id$ is the identity map on $\Bop(\Hch')$,
and $\ket{\Psi}\in \Hch'' \tnsr \Hch'$ is a purification of $\rho$~\cite{schumacher96,barnum00}. 
%%The space $\Hch''$ may be a copy of $\Hch'$ though it is not necessary.
For consistency, we assume that all logarithms appearing in these
entropic quantities are to base $\dmn$ throughout the paper.

This work's main result is the next one.

\begin{theorem} \label{th:main}
Let the basis $\Ebasis=\{ \Ebe_{u} \}_{u\in\cX}=\{ X^i Z^j\}_{(i,j)\in\cX}$
be specified as above.
For a memoryless channel $\cA \sim \{ \sqrt{P(u)} \Ebe_{u} \}_{u\in\cX}$,
where $P$ is a probability distribution on $\cX$, we have
\[
\Capa(\cA) \ge \sup_{\varnin \ge 1}
\max_{ \Hcd \in \rsC_{\varnin}(\Ebasis) }
\frac{I_{\rm c}\big((\dim \Hcd)^{-1} \Pi_{\Hcd},\cA^{\tnsr \varnin}\big)}{\varnin},
\]
where $\Pi_{\Hcd}$ is the projection onto $\Hcd$
and $\rsC_n(\Ebasis)$ is the set of symplectic stabilizer codes,
the precise definition of which %$\rsC_n(\Ebasis)$
will be given in Definition~\ref{def:stab} in Section~\ref{subsec:sc}.
\enth

After proving this, we argue that this bound is actually
the `conditional' capacity of the depolarizing channel on symplectic codes,
which indicates the supremum of information
rates at which symplectic codes work reliably.

\section{Codes Based on Symplectic Geometry \label{ss:sc}}

\subsection{Basics of Symplectic Stabilizer Codes \label{subsec:sc}}

In this section, the framework of symplectic codes is rebuilt
on the theory of geometric algebra~\cite[Chapter~III]{artin}, \cite{grove}.
In designing symplectic codes, we use Weyl's unitary basis~\cite{weyl31},
$\Ebasis=\{ \Ebe_{u} \}_{u\in\cX}$,
which has been specified by (\ref{eq:error_basis0}) and (\ref{eq:error_basis}).
We can regard the index of $\Ebe_{(i,j)}=X^i Z^j$, $(i,j)\in\cX$
%%or $\Ebe'_{(i,j)}=X^i Z^j$, $(i,j)\in\cX$,
as a pair of elements from $\myF=\field_{\dmn}=\bZ/\dmn\bZ$,
the finite field consisting of $\dmn$ elements.
Recall that we have put
$
\Ebe_x = \Ebe_{x_1} \tnsr \dots \tnsr \Ebe_{x_n}
$ 
for $x=(x_1,\dots,x_n)\in (\myF^2)^n$.
We write $\Ebe_{\Icr}$ for %to denote 
$\{ \Ebe_{x} \in \Ebasis_n \mid x\in\Icr \}$
where $\Icr \subset (\myF^2)^n$.
The index $\big((u_1,v_1),\dots,(u_n,v_n)\big)\in (\myF^2)^n$
of a basis element can be regarded as the plain $2n$-dimensional vector
\[
x=(u_1,v_1,\dots,u_n,v_n) \in \myF^{2n}.
\]
%This simplification will frequently be used in what follows.
%
We can equip the vector space $\myF^{2n}$ over $\myF$ with
the standard {\em symplectic bilinear form}\/ (symplectic paring)
which is defined by
\begin{equation}\label{eq:sp}
\syp{x}{y} = \sum_{i=1}^{n} u_i v_i' - v_i u_i'
\end{equation}
for the above $x$ and $y=(u'_1,v'_1,\dots,u'_n,v'_n) \in \myF^{2n}$~\cite{artin,grove}.
For a subspace $\Cso\in \myF^{2n}$, let $\Cso^{\perp}$ be defined by
\[
\Cso^{\perp} = \{ y\in\myF^{2n} \mid \forall x\in\Cso,\ \syp{x}{y}=0 \}.
\]
A subspace $\Cso \in \myF^{2n}$ 
is said to be {\em self-orthogonal}\/ (with respect to the symplectic bilinear form)
if $\Cso\subset \Cso^{\perp}$.

The relation (\ref{eq:XZomega})
implies the following two important properties of $\Ebasis_{n}$
(see, e.g., \cite{weyl31,AshikhminKnill00}):
\begin{equation}\label{eq:propertyEB}
\Ebe_{x} \Ebe_{y} = \zeta_{xy} \Ebe_{x+y}
\end{equation}
for some constants $\zeta_{xy}$ with $|\zeta_{xy}|=1$,
$x,y\in\myF^{2n}$, and
\begin{equation}\label{eq:commutative}
 \Ebe_{x} \Ebe_{y} = \omega^{\syp{x}{y}} \Ebe_{y} \Ebe_{x}.
\end{equation}
The latter implies that
{\em $\syp{x}{y}=0$ if and only if $\Ebe_{x}$ and $\Ebe_{y}$ commute.}

The statement of
the following lemma can be found in
Gottesman~\cite[Section~3.2]{gottesmanPhD}, \cite{gottesman99}.
A simple constructive proof based on
the very basics of symplectic geometry~\cite{artin,grove}
is given in Appendix~\ref{ss:proof4hp}.
%\footnote{The notations $\dim$ and $\mymathsf{span}$ are applied to both 
%spaces or sets of vectors over complex numbers and those over $\myF$.}

\begin{lemma}\label{lem:hyperbolic_plane}
Let $\Cso$ be a self-orthogonal subspace with %ver2
$\dim \Cso = n-k$ and $\Cso= \spn \{ g_1, \dots , g_{n-k} \}$. 
Then, we can find vectors $g_{n-k+1},\dots,g_n$ and $h_1,\dots,h_n$
such that 
\begin{equation}\label{eq:hyperbolic_plane}
\begin{array}{lll}
\syp{g_i}{h_j} &=& \delta_{ij}, \\
\syp{g_i}{g_j} &=& 0,\\
\syp{h_i}{h_j} &=& 0
\end{array}
\end{equation}
for $i,j=1,\dots,n$, where $\delta_{ij}$ is the Kronecker delta.
\enlem

{\em Remark.}\/
In Gottesman's dissertation~\cite{gottesmanPhD}, 
$\Ebe'_{g_i}$ and $\Ebe'_{h_i}$,
$i=n-k+1,\dots,n$ (see Section~\ref{subsec:remark4sc}),
appear as $\Bar{Z}_i$ and $\Bar{X}_i$, respectively,
with examples of them for a number of symplectic stabilizer codes.
\enremark

A pair of
linearly independent vectors $(g,h)$ with $\syp{g}{h}=1$ is called
a {\em hyperbolic pair}\/, and it is known that
a space with a nondegenerate symplectic form, such as the one
defined by (\ref{eq:sp}),
can be decomposed into an orthogonal sum of the form 
\[
\spn \{w_1,z_1\} \perp \dots \perp \spn \{w_n,z_n\}
\]
in such a way that
$(w_i,z_i)$, $i=1,\dots,n$, are hyperbolic pairs~\cite{artin}.
Following Artin~\cite{artin}, we have referred and will refer to
the direct sum of $U_1,\dots,U_n$ as the orthogonal sum of spaces $U_1,\dots,U_n$
if $U_1,\dots,U_n$ are orthogonal. %mutually 
The three equations in the above lemma say
that $\myF^{2n}$ is the orthogonal sum of $\spn \{ g_i,h_i \}$,
$i=1,\dots,n$.
In the present case with the bilinear form in (\ref{eq:sp}),
the simplest example of such a decomposition of the space $\myF^{2n}$
is $\spn \{ e_1,e_2 \} \perp \dots \perp \spn \{ e_{2n-1},e_{2n} \}$,
where $\{ e_i \}_{1\le i \le 2n}$ is the standard basis of $\myF^{2n}$
that consists
of $e_i=(\delta_{ij})_{ 0 \le j \le 2n}\in\myF^{2n}$, $1\le i \le 2n$.

{\em For the remainder of this section, we fix an arbitrary self-orthogonal 
subspace $\Cso$ with $\dim \Cso = n-k$ as in Lemma~\ref{lem:hyperbolic_plane}
and such hyperbolic pairs $(g_1,h_1), \dots, (g_{n},h_{n})$ as just
constructed.}\/
We define
a linear map $\coch: \myF^{2\varnin} \to \myF^{2\varnin}$, %isometry
which depends on the basis $\{ g_1,h_1,\dots,g_{n},h_{n} \}$,
%(w_1,z_1,\dots,w_{\varnin},z_{\varnin}) \mapsto (x_1,\dots,x_{2\varnin}$ 
by
\begin{equation}\label{eq:coch}
\coch ( x ) =
(\tvarz_1,\tvarw_1,\dots,\tvarz_{\varnin},\tvarw_{\varnin})
%= \coch \left( \sum_{i=1}^{\varnin} w_{i} g_i + z_{i} h_i \right) \in \myF^{2\varnin}.
\end{equation}
for a vector $x \in \myF^{2\varnin}$ expanded into
\begin{equation}\label{eq:expand_x}
 x = \sum_{i=1}^{n} (w_i g_i+ z_i h_i).
\end{equation}
The $j$-th coordinate of $\coch(x)$ is denoted by $\cochith{x}{j}$.
In other words, we define $\coch_{j}$ by
\begin{equation}\label{eq:coch_j}
\begin{array}{lll}
{\dss \coch_{2k-1}\left( \sum_{i=1}^{\varnin} (x'_{2i-1} g_i + x'_{2i} h_i) \right)}
&=& x'_{2k-1},\\
{\dss \coch_{2k}\left( \sum_{i=1}^{\varnin} (x'_{2i-1} g_i + x'_{2i} h_i) \right)}
&=& x'_{2k}
\end{array}
\end{equation}
for $k=1,\dots,n$.
For $z=(z_1,\dots,z_m)\in\myF^m$, $1 \le m \le n$, we write
\begin{equation}\label{eq:only_h}
\Ebeh{z}=\prod_{i=1}^{m} (\Ebe_{h_i})^{z_i} %ver2
%\Ebeh{z}=\prod_{i=1}^{m} \Ebe_{h_i}^{z_i}
\end{equation}
where the product on the left-hand side is unambiguous because
$(\Ebe_{h_i})^{z_i}$, $i=1,\dots,m$, commute with each other.
Note that by (\ref{eq:propertyEB}), 
$\Ebeh{z}$ and $\Ebe_{x}$, where $x={\dss \sum_{i=1}^{m} z_i h_i}$,
are the same up to a phase factor.
Similarly, for $w=(w_1,\dots,w_m)\in\myF^m$, $1 \le m \le n$,
we write
\begin{equation}\label{eq:only_g}
\Ebeg{w}=\prod_{i=1}^{m} (\Ebe_{g_i})^{w_i}. %ver2
%\Ebeg{w}=\prod_{i=1}^{m} \Ebe_{g_i}^{w_i}.
\end{equation}
We have seen that any basis $\{ g_1,\dots,g_{n-k} \}$
of a self-orthogonal space can be extended to $\{ g_1,\dots,g_n \}$
in such a way that $\spn \{ g_1,\dots,g_n \}$ is self-orthogonal.
Since $\Ebe_{g_i}$, $i=1,\dots,n$, commute with each other
by (\ref{eq:commutative}),
we can find a basis of $\Bop(\Hch)$ on which $\Ebe_{g_i}$
are simultaneously diagonalized in matrix forms
%see a textbook on quantum mechanics, 
(e.g., \cite{takeuchi}).
Hence, we can find an $n$-tuple of scalars
$( \mu_{i} )_{1\le i\le n}$ for which the space consisting of $\psi$ with
\begin{equation}\label{eq:eigenspace}
\Ebe_{g_i} \psi = \mu_i \psi, \quad i=1,\dots,n,
\end{equation}
is not empty. We call a nonzero vector (respectively, the set of vectors) 
satisfying (\ref{eq:eigenspace})
an eigenvector (respectively, the eigenspace) 
of $\{ \Ebe_{g_i} \}_{1\le i\le n}$ 
with eigenvalue list $( \mu_{i} )_{1\le i\le n}$.
Take a normalized vector 
$\ket{\ghb{0,\dots,0}}$ from this eigenspace, 
where the label $(0,\dots,0)$ belongs to $\myF^n$.
Applying an operator $\Ebe_{x}$ on both side of (\ref{eq:eigenspace})
from left
and using (\ref{eq:commutative}) as well as the symplectic property
\[
\syp{x}{y}=-\syp{y}{x},
\]
we have
\begin{gather}
\Ebe_{x} \Ebe_{g_i} \psi = \mu_i \Ebe_{x} \psi \notag\\
\longleftrightarrow \quad \Ebe_{g_i} \Ebe_{x} \psi = \mu_i \omega^{\syp{g_i}{x}}
\Ebe_{x} \psi. \label{eq:sc_property}
\end{gather}
This means that $\Ebe_{x}\psi$ is an eigenvector with
eigenvalue list $( \mu_{i} \omega^{\syp{g_i}{x}} )_{1\le i\le n}$.
If we expand $x$ %in terms of the basis $g_1,h_1,\dots$
as in (\ref{eq:expand_x}),
then we have $\syp{g_i}{x} = z_i$, $i=1,\dots,n$, and hence
there are, at least, $\dmn^n$ possible eigenvalue lists
for $\{ \Ebe_{g_i} \}_{1\le i \le n}$.
%Note that for a fixed $n$-tuple $(s_i) \in \myF^n$,
%the set $\{ x \mid \syp{g_i}{x}=s_i \}$ is a coset of
%$\Cso_{\rm max}$ in $\myF^{n}$ with a representative $\sum_{i} s_i h_i$.
%and there is a one-to-one correspondence between
However, for any pair of distinct eigenvalue lists, the corresponding
eigenspaces of $\{ \Ebe_{g_i} \}_{1 \le i \le n}$ are orthogonal,
and hence there are no more eigenvalue lists.
Thus, we have an orthonormal basis
$\{\ket{\ghb{s_1,\dots,s_n}}\}_{(s_1,\dots,s_n)\in\myF^n}$ defined by
\begin{equation}\label{eq:basis4Hn}
\ket{\ghb{s_1,\dots,s_n}} = \Ebeh{s} \ket{\ghb{0,\dots,0}},\quad
\mbox{where} \quad {\dss s=(s_1,\dots,s_n)}.
\end{equation}
Note that the basis
$\{\ket{\ghb{s_1,\dots,s_n}}\}_{(s_1,\dots,s_n)\in\myF^n}$
depends on $(g_i,h_i)$, $i=1,\dots,n$, as well as $(\mu_i)_{1\le i \le n}$.

Now we are ready to see %describe
the principle of symplectic codes.

\begin{lemma}\cite{crss97,crss98,gottesman96}. \label{lem:coset_leaders}
Let a subspace $\Cso\subset\myF^{2n}$ satisfy 
\begin{equation}\label{eq:self-orth}
\Cso\subset \Cso^{\perp} \quad \mbox{and} \quad \dim \Cso = n-k.
\end{equation}
In addition, let $\Icr_0 \subset \myF^{2n}$ 
be a set satisfying
\begin{equation} %\label{eq:coset_leaders} %number and label not used
\forall x,y\in \Icr_0,\ [ \, y-x \in \Cso^{\perp} \Rightarrow x=y \, ],
\end{equation}
and put
\[
 \Icr=\Icr_0+\Cso=\{ z+w \mid z\in\Icr_0,w\in\Cso \}.
\]
Then, the $\dmn^{k}$-dimensional subspaces of the form
\begin{equation}\label{eq:codespace}
\{ \psi \in \Hch^{\tnsr n} \mid \forall M\in\Ebe_{\Cso},\ M \psi =  \tau(M) \psi \},
\end{equation}
where $\tau(M)$ are eigenvalues of $M\in \Ebe_{\Cso}$,
are $\Ebe_{\Icr}$-correcting codes.
\enlem

In fact, the subspace %NOTE LATER to be changed for LAYOUT
\begin{equation}\label{eq:codespace2}
\Hcd^{(s)} = \spn \{ \ket{\ghb{\varsp_1,\dots,\varsp_{n-k},\varsp_{n-k+1},\dots,\varsp_n}} \mid (\varsp_{n-k+1},\dots,\varsp_n) \in \myF^k \}
\end{equation}
with a fixed $(n-k)$-tuple $s=(\varsp_1,\dots,\varsp_{n-k})\in\myF^{n-k}$ 
is such a quantum code.
The equivalence of (\ref{eq:codespace}) and (\ref{eq:codespace2})
follows from (\ref{eq:propertyEB}).
Since there are $\dmn^{n-k}$ possible choices for $(\varsp_1,\dots,\varsp_{n-k})$,
we have $\dmn^{n-k}$ codes.
The term {\em codes}\/ is applied to both a self-orthogonal subspace
$\Cso \subset \myF^{2n}$, and quantum codes $\Hcd^{(s)}$
associated with $\Cso$, 
which we will call {\em symplectic (stabilizer) codes with
stabilizer}\/ $\Ebe_{\Cso}$.
Since $\Cso^{\perp}$ is spanned by $g_1,\dots,g_n$ and $h_{n-k+1},\dots,h_n$,
any coset of $\Cso^{\perp}$ in $\myF^{2n}$ is of the form
\begin{equation}\label{eq:coset}
\Big\{   \sum_{i=1}^{n} (w_i g_i+ z_i h_i) 
\mid z_i = s_i,\, i=1,\dots,n-k \Big\}
=\{ x \mid \syp{g_i}{x} = s_i,\, i=1,\dots,n-k  \}
\end{equation}
with some $(n-k)$-tuple $s=(s_1,\dots,s_{n-k})$.
In terms of $\coch_j$ defined by (\ref{eq:coch_j}),
the coset in (\ref{eq:coset}) can be rewritten as
\begin{equation}\label{eq:synd}
\{ x \mid \cochith{x}{2i}=s_i,\, i=1,\dots, n-k \}.
\end{equation}
The set of cosets of $\Cso^{\perp}$ and 
$\{ \Ebe_{\varrc} \Hcd^{(0)} \mid \varrc \in \Icr_0 \}$,
where $\Ebe_{\varrc} \Hcd^{(0)}$ denotes $\{ \Ebe_{\varrc}\psi \mid \psi\in\Hcd^{(0)} \}$ with $0=(0,\dots,0)\in\myF^{n-k}$,
are in a one-to-one correspondence
when $\Icr_0$ is a transversal (a complete set of coset representatives),
i.e., when $\crd{\Icr_0}=\dmn^{n-k}$.
%, in view of (\ref{eq:sc_property}), (\ref{eq:basis4Hn}) and (\ref{eq:codespace2}). %?
In fact, for any vector $x$ in the coset in (\ref{eq:coset})
or (\ref{eq:synd}), we have (cf.\ Section~\ref{subsec:te} below)
\begin{equation}\label{eq:synd2}
\Hcd^{(s)}= \Ebe_{\varrc} \Hcd^{(0)}.
\end{equation}
The $(n-k)$-tuple $(s_i)_{1 \le i \le n-k}$ is called a syndrome on the analogy
with classical linear codes. %by the analogy?

To show that the subspace, say $\Hcd$, 
in (\ref{eq:codespace}) or (\ref{eq:codespace2})
is really $\Ebe_{\Icr}$-correcting, we may use 
Theorem III.2 of Knill and Laflamme~\cite{KnillLaflamme97}.
Alternatively, we can directly check the error-correcting capability
using the recovery operator $\cR$ defined by 
\begin{equation} \label{eq:recover}
\cR \sim \{ \Proj_{\rm res} \} \cup \{ \Ebe_{\varrc}^{\dagger} \Proj_{\varrc}
\}_{\varrc\in\Icr_0},
\end{equation}
where $\Proj_{\varrc}$ is the projection onto
$\Ebe_{\varrc} \Hcd= \{ \Ebe_{\varrc}\psi \mid \psi\in\Hcd \}$,
and $\Proj_{\rm res}$ is the projection onto the orthogonal complement
of $\bigoplus_{\varrc\in\Icr_0} \Ebe_{\varrc} \Hcd$ in $\Hch^{\tnsr n}$.

To be specific about which class of codes we are treating,
we define the next.

\begin{definition}\label{def:stab}
We define $\rsC_n(\Ebasis)$, $n \ge 1$,
to be the set of all symplectic stabilizer codes with stabilizer $\Ebe_{\Cso}$,
i.e., all $\dmn^k$-dimensional
subspaces of $\Hch^{\tnsr n}$ of the form (\ref{eq:codespace})
or (\ref{eq:codespace2}),
with some subspace $L\subset\myF^{2n}$ satisfying (\ref{eq:self-orth})
for some $k$, $0 \le k \le n$.
\closedef

\subsection{Tracing Errors \label{subsec:te}}

Viewing index vectors in terms of the basis $\{ g_1,h_1,\dots,g_{n},h_{n} \}$
is also useful to trace the action of an error in $\Ebasis_n$
on a state in the code space. The view introduced in this subsection, 
as well as that in the next one,
will underlie the proof of the main result to be given later. %below
Let us consider the code $\Hcd^{(s)}$ in (\ref{eq:codespace2})
assuming $\varsp_1=\dots=\varsp_{n-k}=0$, which loses no generality
since $(\mu_i)_{1 \le i \le n}$ is arbitrary.

Suppose an error $\Ebe_{x}$ has occurred on
a state $\rho$ whose range (image, or support) is contained
in the code space $\Hcd^{(s)}$.
We expand $x$ as in (\ref{eq:expand_x}) and put
\begin{equation}
\begin{array}{lll}
t &=& (w_1,\dots,w_{n-k}), \\
\vartpr &=& ( z_1,\dots,z_{n-k}), \\
u_i &=& z_{i+n-k} , \quad i=1,\dots,k, \\
u'_i &=& w_{i+n-k} , \quad i=1,\dots,k,\\
u &=& (u_1,\dots,u_k), \\ 
u' &=& (u'_1,\dots,u'_k).
\end{array}\label{eq:te0}
\end{equation}
Then, for the purpose of analysis, 
we interpret the action of $\Ebe_{x}$ as follows:
First, $\Ebeg{t}$ occurred to make no change on $\rho$,
second, $\Ebeh{\vartpr}$ occurred to change $\rho$, 
which is a linear combination of 
\[
\ket{\ghb{0,\dots,0,\varb_{1},\dots,\varb_k}} \bra{\ghb{0,\dots,0,\vard_{1},\dots,\vard_k}}, \quad (b_1,\dots,b_k),(b'_1,\dots,b'_k)\in\myF^{k},
\]
into the linear combination $\rho'$ of
\[
\ket{\ghb{z_1,\dots,z_{n-k},\varb_{1},\dots,\varb_k}} \bra{\ghb{z_1,\dots,z_{n-k},\vard_{1},\dots,\vard_k}},\quad (b_1,\dots,b_k),(b'_1,\dots,b'_k)\in\myF^{k},
\]
with the same coefficients by (\ref{eq:basis4Hn}),
and finally,
$\Bar{X}_{u}\Bar{Z}_{u'}$ occurred to act on $\rho'$ as 
$\Bar{X}_{u} \Bar{Z}_{u'}\rho' \Bar{Z}_{u'}^{\dagger}\Bar{X}_{u}^{\dagger}$,
where the actions of $\Bar{X}_{u}$ and $\Bar{Z}_{u'}$,
$u,u'\in\myF^k$,
are defined by
\[
\Bar{X}_{u}\ket{\ghb{z_1,\dots,z_{n-k},\varb_{1},\dots,\varb_k}}
=\ket{\ghb{z_1,\dots,z_{n-k},\varb_{1}+u_1,\dots,\varb_k+u_k}}
\]
and
\[
\Bar{Z}_{u'}\ket{\ghb{z_1,\dots,z_{n-k},\varb_{1},\dots,\varb_k}}
=\prod_{i=1}^{k}\omega^{u'_i \varb_i}\ket{\ghb{z_1,\dots,z_{n-k},\varb_{1},\dots,\varb_k}}
\]
for $(z_1,\dots,z_{n-k})\in\myF^{n-k}$, $(b_1,\dots,b_k)\in\myF^{k}$.
In other words, we have the next.

\begin{lemma}\label{lem:te}
Let us given $\Cso$ and $\{ g_1,h_1,\dots,g_{n},h_{n} \}$ as above.
Then, we have
\begin{eqnarray}	
\Ebe_{x}\rho\Ebe_{x}^{\dagger} &=&
\Bar{X}_{u} \Bar{Z}_{u'} \Ebeh{\vartpr} \Ebeg{t} \rho \Ebeg{t}\mbox{}^{\dagger}  
\Ebeh{\vartpr}\mbox{}^{\dagger} \Bar{Z}_{u'}^{\dagger} \Bar{X}_{u}^{\dagger} \notag\\
 &=&
\Bar{X}_{u} \Bar{Z}_{u'} \Ebeh{\vartpr} \rho 
\Ebeh{\vartpr}\mbox{}^{\dagger} \Bar{Z}_{u'}^{\dagger} \Bar{X}_{u}^{\dagger}\label{eq:te}
\end{eqnarray}
for any operator $\rho$ such that the ranges of $\rho$ and $\rho^{\dagger}$
are contained in the code space in %ver3
(\ref{eq:codespace2}) and for any $x\in\myF^{2n}$, where $t,\vartpr,u,u'$
are determined from $x$ through (\ref{eq:te0}).
\end{lemma}

This is clear from (\ref{eq:propertyEB}) and
\[
x=\sum_{i=1}^{n-k} w_i g_i +\sum_{i=1}^{n-k} z_i h_i + \sum_{i=1}^{k} u_i h_{i+n-k}+\sum_{i=1}^{k} u'_i g_{i+n-k}
\]
for $\rho=\Pi_{\Hcd^{(0)}}$. %$\rho=\dmn^{-k}\Pi_{\Hcd^{(0)}$.
For a general operator $\rho$, we should consider phase factors
as is done in Appendix~\ref{ss:proof4te}.
Observe that the action of $\Bar{X}_u\Bar{Z}_{u'}$ is
similar to that of $\Ebe_{(v_1,v'_1,\dots,v_n,v'_n)} =
(X^{v_1} \tnsr \dots \tnsr X^{v_n})( Z^{v'_1}\tnsr \dots \tnsr Z^{v'_n})
\in\Ebasis_n$ on states 
$\ket{0\dots 0},\dots,\ket{1\dots 1} \in \Hch^{\tnsr n}$.

\subsection{Coset Arrays and Probability Arrays \label{ss:ca}}

To understand the action of errors in $\Ebasis_n$ on symplectic codes
associated with the self-orthogonal subspace $\Cso\in\myF^{2n}$,
it is helpful to consider cosets of $\Cso$.
%%, and such a view will underlie the arguments to be given below.
Since $\dim \Cso=n-k$ implies $\dim \Cso^{\perp}=n+k$
(Lemma~\ref{lem:hyperbolic_plane}),
we have $\dmn^{n-k}$ cosets of
$\Cso^{\perp}$ in $\myF^{2n}$, and each coset is a union of
$\dmn^{2k}$ cosets of $\Cso$ in $\myF^{2n}$.
To grasp the situation, we write down an array of cosets,
which we will call a {\em coset array}\/ of $\Cso$, as follows:
\begin{equation}\label{eq:carray1}
\begin{array}{rrrr}
y_0+x_0+\Cso & y_0+x_1 +\Cso  & \cdots & y_0+x_{\varDK-1} +\Cso \\
y_1+x_0+\Cso & y_1 + x_1 +\Cso  & \cdots & y_1+x_{\varDK-1} +\Cso \\
\vdots\ & \vdots\ & & \vdots\ \\
y_{\varNK-1}+x_0+\Cso & y_{\varNK-1}+ x_1 +\Cso  & \cdots & y_{\varNK-1}+x_{\varDK-1} +\Cso
\end{array}
\end{equation}
where $\varDK=\dmn^{2k}$, $\varNK=\dmn^{n-k}$,
$\{ x_i \}$ is a transversal of the cosets of $\Cso$ in $\Cso^{\perp}$,
and $\{ y_i \} $ is that of the cosets of $\Cso^{\perp}$ in $\myF^{2n}$.
In the array, each entry is a coset of $\Cso$ in $\myF^{2n}$,
and each row form a coset of $\Cso^{\perp}$ in $\myF^{2n}$.
This array, which has
appeared in Fig.~1 of DiVincenzo {\em et al.}\/~\cite{dss98}
in a different configuration, resembles standard arrays
often used in classical coding theory~\cite{slepian56,ptrsn} 
though they differ in that elements of standard arrays
are vectors rather than cosets. %ver2
%This array is useful to understand the decoding principle.
%From each row, we chose a coset leader to form $\Icr_0$.

%In Section~\ref{subsec:sc}, 
We have already seen that
cosets of $\Cso^{\perp}$ can be labeled with $s\in\myF^{n-k}$
as in (\ref{eq:coset}) or (\ref{eq:synd}). Furthermore,
using hyperbolic pairs $(g_i,h_i)$, $i=1,\dots,n$, 
%which form a basis of $\myF^{2n}$, 
as in Lemma~\ref{lem:hyperbolic_plane},
we can label cosets of $\Cso$ in $\Cso^{\perp}$
by $(u_1,u_1',\dots,u_k,u'_k)\in\myF^{2k}$.
In fact, since $(g_i,h_i)$, $i=n-k+1,\dots,n$, together with
the basis elements $g_i$, $i=1,\dots,n-k$, of $\Cso$, form a basis
of $\Cso^{\perp}$,
each coset of $\Cso$ in $\Cso^{\perp}$
can be written in the form 
\[
\{ x\in\myF^{2n} \mid \cochith{x}{2i-1}=u_{i-n+k},\,
\cochith{x}{2i}=u'_{i-n+k}\ {\rm for} \
n-k+1 \le i \le n \}
\]
with some $(u_1,u_1',\dots,u_k,u'_k)\in\myF^{2k}$.
As a result, each coset of $\Cso$ in $\myF^{2n}$
can be specified by some $(s_1,\dots,s_{n-k})\in\myF^{n-k}$ and
$(u_1,u_1',\dots,u_k,u'_k)\in\myF^{2k}$
as the set that consists of the vectors $x\in\myF^{2n}$ satisfying
\[
\cochith{x}{2i}=s_i\,\, {\rm for}\,\, 1\le i \le \varnin-\varkin \quad {\rm and} \quad
\cochith{x}{2i-1}=u_{i-n+k},\
\cochith{x}{2i}=u'_{i-n+k}\,\, {\rm for}\,\,
n-k+1 \le i \le n.
\]
Keeping this labeling in mind,
we will introduce another important quantities,
in terms of which our bound on the capacity will be described.
Given a channel
$\cA \sim \{ \sqrt{P(u)} \Ebe_{u} \}_{u\in\cX}$,
we define a probability distribution $P_{\Cso}$ by
\begin{equation}\label{eq:Pin}
\Pin\big((s,\varupr)\big) = \sum_{x:\,\, \cochith{x}{2i}=s_i\,\, {\rm for}\,\, 1\le i \le
\varnin-\varkin\,\, {\rm and} \,\, \cochith{x}{2i-1}=u_{i-n+k},\, 
\cochith{x}{2i}=u'_{i-n+k}\,\, {\rm for}\,\,
\varnin-\varkin +1 \le i \le \varnin} P^{\varnin}(x),
\end{equation}
where $s=(s_1,\dots,s_{n-k})\in\myF^{n-k}, \varupr=(u_1,u_1',\dots,u_{k},u_k')\in\myF^{2k}$.
Now, arrange $\Pin(s,\varupr)=\Pin\big((s,\varupr)\big)$ into the array of probabilities
\begin{equation} \label{eq:carray2}
\begin{array}{rrrr}
\Pin(0_{n-k},0_{2k}) & \Pin(0_{n-k},0\dots 01)  & \cdots & \Pin(0_{n-k},11 \dots 1) \\
\Pin(0\dots 01,0_{2k}) & \Pin(0\dots 01,0\dots 01)  & \cdots & \Pin(0\dots 01,11 \dots 1) \\
\vdots\ & \vdots\ & & \vdots\ \\
\Pin(11 \dots 1,0_{2k}) & \Pin(11\dots 1,0\dots 01)  & \cdots & \Pin(11\dots 1,11 \dots 1) 
\end{array} %,
\end{equation}
where $0_m$ denotes the zero vector in $\myF^m$ and an $m$-tuple
$(b_1,\dots,b_m)\in\myF^m$ is simply written as $b_1\dots b_m$.
We have assumed here $\dmn=2$ in order that it may not look complicated,
the general description being obvious.
We will call this a {\em probability array}\/ of $\Cso$.
Each probability $\Pin(s,\varupr)$ is the probability of 
the corresponding entry in (\ref{eq:carray1}) if the coset representatives
$x_i$ and $y_j$ are chosen accordingly. %appropriately.
Note that the index $\vartpr$ in Section~\ref{subsec:te} corresponds to
the row index $s$ in the probability array, and $u_i$, $u_i'$
are used for the column index.
We remark that $\Pin$ depends on the choice of hyperbolic pairs
$(g_i,h_i)$, $i=1,\dots,n$, but the probability array of $\Cso$
is unique up to permutations of rows and columns.

\subsection{Decoding Symplectic Stabilizer Codes \label{ss:decode}}

Coset arrays are useful to understand the decoding principle
of symplectic stabilizer codes.
Let us given a code with stabilizer $\Ebe_{\Cso}$.
As explained in Section~\ref{subsec:sc},
once we specify $\Icr_0$, (a subset of)
a transversal of the cosets of $\Cso^{\perp}$ in $\myF^{2n}$, 
the recovery operator of symplectic codes is determined from $\Icr_0$
as in (\ref{eq:recover}) in such a way that
the code can correct errors in $\Ebe_{\Icr}$, where $\Icr=\Icr_0+\Cso$.
The set $\Icr=\Icr_0+\Cso$ is a union of some cosets of $\Cso$,
and in view of (\ref{eq:carray1}), each row of the coset arrays
has exactly one coset (or none) which is a constituent of $\Icr$.
Thus, the design of a decoder of symplectic codes with stabilizer $\Ebe_{\Cso}$
is accomplished by choosing a coset from each row of the array.
When the code in Lemma~\ref{lem:coset_leaders} is
used on a memoryless channel 
$\cA \sim \{ \sqrt{P(u)} \Ebe_{u} \}_{u\in\cX}$,
a natural choice for such a coset in each row may be
one that has the largest value of $P_{\Cso}$ in the row,
since it is analogous to maximum likelihood decoding,
which is an optimum strategy for classical coding.
%(see, e.g., Slepian~\cite{slepian56}).
Our codes to be proven to have the desired performance are
concatenated codes, and our choice for $\Icr_0$ 
will turn out to be more technical exploiting the structure of
concatenated codes.

\subsection{Remarks on Symplectic Stabilizer Codes \label{subsec:remark4sc}}

%Here are some remarks.
% about treatments on symplectic stabilizer codes.
%that can be found in the literature.
When $\dmn=2$, often used is
the slightly different basis $\{ \Ebe'_{(i,j)} \}_{(i,j)\in\cX}$
the elements of which are defined by
$\Ebe'_{(i,j)} = X^i Z^j=\Ebe_{(i,j)}$ for $(0,0),(0,1),(1,0)\in \cX$ and
$\Ebe'_{(1,1)} = \imu X Z = \imu \Ebe_{(1,1)}$~\cite{crss98,gottesmanPhD}.
The arguments below all work if $\Ebe'$ is used instead of $\Ebe$ for $\dmn=2$.

As already mentioned, the recovery operator for an $N_{\Icr}$-correcting
code in Lemma~\ref{lem:coset_leaders} is given by
$\cR \sim \{ \Ebe_{x}^{\dagger} \Proj_x \}_{x\in\Icr_0}$,
where we assume $\crd{\Icr_0}=\dmn^{n-k}$
%$\Icr_0$ is a transversal 
for simplicity.
A physical meaning of this recovery process is simple: It can be described as
the orthogonal measurement $\{ \Proj_x \}_{x\in\Icr_0}$
followed by the unitary operation $\Ebe_{x}^{\dagger}$, which
is chosen accordingly to the measurement result $x$.
The measurement $\{ \Proj_x \}_{x\in\Icr_0}$ is realized by
the observables $\Ebe'_{g_i}$, $i=1,\dots,n-k$,
when $\dmn=2$ (and similar self-adjoint operators that correspond
to $\Ebe_{g}$, $i=1,\dots,n-k$, for $\dmn>2$).
In this case, the syndrome
%, which is in the one-to-one correspondence with $x\in\Icr_0$, 
is obtained as a measurement result.

\section{Concatenated Codes \label{ss:cat}}

%\subsection{Structure of Concatenated Codes}

The very first quantum code discovered by Shor~\cite{shor95}
is an example of a concatenated code.
The idea of the following general code construction
%which allows code parameters to be arbitrary, 
can be found in \cite[Section~3.5]{gottesmanPhD}.
Let $\Cin \subset \myF^{2\varnin}$ and $\Cout\subset \myF^{2\varkin \varN}$ 
be self-orthogonal codes with $\dim \Cin = \varnin
- \varkin$ and $\dim \Cout = \varkin \varN-\varK$. 
Let $\{ g_1, \dots, g_{\varnin-\varkin} \}$ and $\{ g'_1, \dots, g'_{\varkin \varN-\varK}
\}$ be bases of $\Cin$ and $\Cout$, respectively.
Let $\{ g_1, \dots, g_{\varnin-\varkin} \}$ be supplemented by
$g_{\varnin-\varkin+1},\dots,g_{\varnin}$ and $h_1,\dots,h_{\varnin}$ to form
a basis of the property in Lemma~\ref{lem:hyperbolic_plane}.
%$(g_i,h_i)$ become hyperbolic pairs of the property
We will construct a new self-orthogonal code of length $ 2 \varnin \varN$ from 
$\Cin$ and $\Cout$.

For any vector $x=(x_1,\dots,x_{2\varnin})\in\myF^{2\varnin}$,  
let $x^{(j)}$ denote the vector 
$(0,\dots,0, x, 0,\dots,0)\in\myF^{2\varnin\varN}$, where we have divided the $2 \varnin \varN$
coordinates into $\varN$ blocks of length $2 \varnin$ and
$x$ appears at the $j$-th block. 
Next, for any 
$x=(u_{1,1},u'_{1,1},\dots,u_{1,\varkin},u'_{1,\varkin},\dots,
u_{\varN,1},u'_{\varN,1},\dots,u_{\varN,\varkin},u'_{\varN,\varkin})
\in\myF^{2\varkin \varN }$,
let us denote by $\Bar{x}\in\myF^{2 \varnin\varN}$ the vector specified by
\begin{equation} \label{eq:bar_g}
 \Bar{x} = \sum_{j=1}^{\varN}\sum_{m=1}^{\varkin} 
u_{j,m} g_{\varnin-\varkin+m}^{(j)} + u'_{j,m} h_{\varnin-\varkin+m}^{(j)}.
\end{equation}
Especially, we apply this map to $g'_i$ to obtain $\Bar{g'_i}$,
$i=1,\dots,\varkin \varN-\varK$.
Note that the map that sends $x$ to $\Bar{x}$ preserves
the symplectic inner product.
This is because
%$\myF^{2\varkin \varN}= \spn \{ e_1,e_2 \} \perp \dots \perp \spn \{
%e_{2n-1},e_{2n} \}$, 
mutually orthogonal hyperbolic pairs $(e_{2i-1},e_{2i})$,
$i=1,\dots,\varkin \varN$,
are mapped to mutually orthogonal hyperbolic pairs 
$(g_{\varnin-\varkin+m}^{(j)},h_{\varnin-\varkin+m}^{(j)})$, $j=1,\dots,\varN$,
$m=1,\dots,\varkin$, where 
$\{ e_i \}_{1 \le i \le 2\varkin\varN}$ is the standard basis of $\myF^{2\varkin \varN}$.
Clearly, 
\begin{equation*} %\label{eq:Gin}
 \Gin =\{ g_i^{(j)} \mid i=1,\dots,\varnin-\varkin;\, j=1,\dots,\varN \}
\end{equation*}
is a set of mutually orthogonal independent vectors.
Since $\Bar{g'_i}$, $i=1,\dots,\varkin\varN-\varK$, are spanned by 
$g_{\varnin-\varkin+m}^{(j)}$ and $h_{\varnin-\varkin+m}^{(j)}$,
$j=1,\dots,\varN$, $m=1,\dots,\varkin$, 
which are orthogonal to each element of $\Gin$,
we see that 
$\Gin \cup \{ \Bar{g'_i} \mid i=1,\dots,\varkin \varN-\varK\} \subset \myF^{2\varnin \varN}$ 
is a basis of a self-orthogonal code of dimension 
$(\varnin-\varkin)\varN+\varkin \varN-\varK = \varnin \varN-\varK$.
The code over $\myF$ obtained by concatenating two codes 
$\Cin$ and ${\Cout}$ in this way will be
denoted by $\cnc{\Cin}{\Cout}$. Symplectic quantum codes
associated with $\cnc{\Cin}{\Cout}\subset\myF^{2\varnin\varN}$ 
of the above parameters
have information rate $\varK/(\varnin\varN)$.

%{\em Example.}\/
Examples of codes with inner codes
having parameter $\varkin=1$ can be found in the literature~\cite{ss97,dss98,gottesmanPhD}.
For instance, a code with $\varnin=\varN=5$ and $\Cin=\Cout$
was given by Gottesman~\cite[Section~3.5, Table~3.7]{gottesmanPhD}
with a table of $\{ \Ebe'_{g} \mid g\in\cnc{\Cin}{\Cout} \}$.

\section{Proof of Theorem~\ref{th:main} and Remarks \label{ss:proof}}

\subsection{Proof of Theorem~\ref{th:main} \label{sss:proof}}

The theorem can be obtained as a consequence of
the next stronger statement, which is proved in Appendix~\ref{ss:proof2}.

\begin{theorem}\label{th:main2}
Let a function $\Enk{\varnin}{\varkin}$ be defined by
\begin{equation}
\Enk{\varnin}{\varkin}(R,\Pin)
= \min_{P'}
[D(P'||\Pin) + |\varkin-kR-H(\overleftarrow{P'}|\Bar{P'})|^+ ]
\end{equation}
where $|x|^+=\max\{x,0 \}$
and the minimum with respect to $P'$ is taken over all
probability distributions on 
$\myF^{\varnin-\varkin} \times \myF^{2\varkin}$.
Let $R'$ satisfy $0 \le R' \le 1$.
Then, for a memoryless channel 
$\cA \sim \{ \sqrt{P(u)} \Ebe_{u} \}_{u\in\cX}$ and any self-orthogonal
code $\Cin \subset \myF^{2\varnin}$ with dimension $\varnin-\varkin$,
$1 \le \varkin \le \varnin$, we have
\begin{equation} \label{eq:exp_1st}
\limsup_{\varM\to\infty} - \frac{\log_{\dmn}[1-F_{\varM,R'\varM}^{\star}
(\cA^{\tnsr\varM})]}{\varM}
\ge \frac{{\Enk{\varnin}{\varkin}(R'\varnin/k,\Pin)}}{n}
\end{equation}
where $\Pin$ is the probability distribution on
$\myF^{\varnin-\varkin} \times \myF^{2\varkin}$ defined by (\ref{eq:Pin}).
\enth

From the general property of the Kullback-Leibler information $D$
that $D(P||Q) \ge 0$ with equality if and only if $P=Q$,
it follows that $\Enk{\varnin}{\varkin}(R,\Pin)$ is positive
if $ kR < \varkin-H(\overleftarrow{\Pin}|\Bar{\Pin})$.
%if $\varkin-H(\overleftarrow{\Pin}|\Bar{\Pin}) > kR$ ($=nR'$).
Hence, $[\varkin-H(\overleftarrow{\Pin}|\Bar{\Pin})]/\varnin$ is a lower bound on
the capacity of the channel. Thus, the next corollary follows.

\begin{corollary} \label{coro:1}
For the memoryless channel $\cA$, we have
\[
%\Capa(\cA|\{ \rsC_n \}) 
\Capa(\cA)
\ge \sup_{\varnin \ge 1}\, \max_{\Cin} \frac{\varkin-H(\overleftarrow{\Pin}|\Bar{\Pin})}{\varnin},
\]
where the maximum with respect to $\Cin$ is over all $\varkin$
with $1 \le \varkin \le \varnin$ and all %self-orthogonal subspace
$\Cin \subset \myF^{2\varnin}$ with $\dim \Cin=\varnin-\varkin$ and 
$\Cin \subset \Cin^{\perp}$.
\end{corollary}

{\em Remark.}\/
When $\varnin=\varkin$,
a coset array of $\Cso$ consists of a single row, and
$H(\overleftarrow{\Pin}|\Bar{\Pin})$ is to be understood as
$H(\Pin)=\varnin H(P)$.
%where the equality is due to the fact that
%$\Pin$ becomes the product measure $P^{\varnin}$ on $(\myF^2)^{\varnin}$.
In this case, $[k-H(\overleftarrow{\Pin}|\Bar{\Pin})]/n = 1-H(P)$,
which is the known lower bound~\cite{preskillLNbook0,hamada01e,hamada01g}.
\enremark

We also have the next.

\begin{lemma} \label{lem:Pin_Ic}
Let $\cA$ be the one in Theorem~\ref{th:main},
viz., $\cA\sim \{ \sqrt{P(u)} \Ebe_u \}_{u\in\cX}$,
$\Cso$ be a self-orthogonal code satisfying (\ref{eq:self-orth}), and
$\Pin$ be defined by (\ref{eq:Pin}).
Then, we have
\[
\varkin-H(\overleftarrow{\Pin}|\Bar{\Pin}) = I_{\rm c}\big((\dim \Hcd)^{-1}\Pi_{\Hcd},\cA^{\tnsr n}\big),
\]
where $\Hcd$ is a symplectic code
with stabilizer $\Ebe_{\Cso}$.
\enlem

A proof is given in Appendix~\ref{ss:proof4Lemma5}.
Corollary~\ref{coro:1} and Lemma~\ref{lem:Pin_Ic} are extensions of the facts
established by Shor and Smolin~\cite{ss97} and 
DiVincenzo and these authors~\cite{dss98},
who restricted $\Cso$ to those of quantum repetition codes
having parameter $k=1$.

Corollary~\ref{coro:1}, together with Lemma~\ref{lem:Pin_Ic},
establishes Theorem~\ref{th:main}.

\subsection{Remarks on Theorems~\ref{th:main} and \ref{th:main2}
\label{ss:rem_mains}}
%\subsection{Visualizing the Lower Bounds \label{ss:visual}}

The quantity $H(\overleftarrow{\Pin}|\Bar{\Pin})$ appearing
Corollary~\ref{coro:1} can be written %ver2
solely in terms of $\Cso$, which specifies the quantum code,
and $P$, which specifies the channels,
and it does not depend on the choice of hyperbolic pairs $(g_i,h_i)$,
$i=1,\dots,n$, 
since $H(\overleftarrow{\Pin}|\Bar{\Pin})$ is a function
of the array or matrix in (\ref{eq:carray2})
as is mentioned in the proof of Lemma~\ref{lem:Pin_Ic},
Appendix~\ref{ss:proof4Lemma5}, and its value does not
change if we permute rows or columns of the array.
%with the channel $\cA \sim \{ \sqrt{P(u)} \Ebe_u \}_{u\in\cX}$ fixed,
Similarly, $\Enk{\varnin}{\varkin}(R,\Pin)$
do not depend on the choice of
hyperbolic pairs $(g_i,h_i)$, $i=1,\dots,n$.

\subsection{Idea for Proof of Theorem~\ref{th:main2} \label{ss:on_main2}}

%\subsection{Decoding of Concatenated Codes \label{ss:decode_cat}}

The theorem is proved with
a random coding argument similar to those in \cite{hamada01e,hamada01g},
the main difference being in the decoding strategy.
A concatenated code associated with $\cnc{\Cin}{\Cout}$
is a symplectic stabilizer code,
so that we can apply the decoding strategy described in 
Section~\ref{ss:decode} to it.
Especially, minimum entropy decoder employed in \cite{hamada01e,hamada01g}
can be used.
In the proof of Theorem~\ref{th:main2}, however, we modify
this decoding strategy incorporating Shor and Smolin's idea.
Namely, we choose
a vector that minimizes the {\em conditional}\/ entropy of the type of it
in each coset of $\cnc{\Cin}{\Cout}^{\perp}$ in $\myF^{2\varnin\varN}$,
where the conditioning is on
the result of measuring the observables $\Ebe'_{g}$, $g \in\Gin$,
when $\dmn=2$, or similar ones for $\dmn>2$,
which form a part of the syndrome of the concatenated code.
%Details will be given in the proof below.

\section{Conditional Capacity \label{ss:cond_cap}}

\subsection{Conditional Capacity and Upper Bound}

In discussing capacity problems on classical channels, we sometimes 
put restriction on coding schemes.
For example, there are works on the highest
information rate achievable by linear codes~\cite{gabidulin67},
the conditional capacity with cost or power constraints and so on.
In a similar way, we discuss a conditional quantum capacity
in this section.
Suppose for each $n>0$, a set $\rsCgen_{n}$ 
of subspaces of $\Hch^{\tnsr n}$ is given.
We imagine the situation in which
only subspaces belonging to $\rsCgen_{n}$ can be used as codes.

\begin{definition} \label{def:cond_cap}
Let a sequence of code classes $\{ \rsCgen_n \}$ be given, and 
$F_{n,k}^{\star}(\cA^{\tnsr n} | \rsCgen_n )$ denote the
supremum of
$F(\Hcd, \cR\cA^{\tnsr n})$ such that there exists a code 
$(\Hcd, \cR)$ with $\Hcd \in \rsCgen_n$
and $\log_{\dmn} \dim \Hcd \ge k$, where $n>0$ is an integer
while $k$, $0 \le k \le n$, is a real number.
The supremum of nonnegative numbers $R$ satisfying
\[
\limsup_{n\to\infty} F_{n,Rn}^{\star}(\cA^{\tnsr n}|
\rsCgen_n )=1
\]
is called the conditional quantum capacity of $\cA$ on $\{ \rsCgen_n \}$ and
denoted by $\Capa(\cA|\{ \rsCgen_n \})$.
\closedef

Comparing this with Definition~\ref{def:cap}, we see
$\Capa(\cA)=\Capa(\cA| \{\rsCgen_n \})$
when we put no restriction on coding schemes,
i.e., when $\rsCgen_n$ is the set of all subspaces of $\Hch^{\tnsr n}$
for each $n>0$.

%\subsection{Upper Bound on Conditional Capacity}

We have an upper bound on the conditional capacity,
a proof of which is given in Appendix~\ref{ss:proof4ub}.

\begin{lemma}\label{lem:ub4cond_cap}
Let a sequence of code classes $\{ \rsCgen_n \}$ be given. Then,
\begin{equation}\label{eq:upper_bound}
\Capa(\cA|\{ \rsCgen_n \}) \le 
\limsup_{\varnin \to \infty} \sup_{\Hcd \in \rsCgen_{\varnin} }
\frac{ I_{\rm c}\big((\dim \Hcd)^{-1} \Pi_{\Hcd}, \cA^{\tnsr \varnin}\big)}{\varnin},
\end{equation}
where $\Pi_{\Hcd}$ is the projection onto $\Hcd$.
\enlem

\subsection{Conditional Capacity of the Depolarizing Channel on Stabilizer Codes}

In this subsection, we will see
that the lower bound on the capacity
obtained in the previous section is, in fact,
a lower bound on the conditional capacity $\Capa(\cA | \{ \rsC_n \})$
of the depolarizing channel,
where $\rsC_n$ is the set of all symplectic stabilizer codes.
To be precise, we put
%we first extend $\rsC_n(\Ebasis)$ in Definition~\ref{def:stab}
%to the wider class. 
\[
\rsC_n=\bigcup_{\Ebasis} \rsC_n(\Ebasis),
\]
where 
$\rsC_n(\Ebasis)$ is defined in Definition~\ref{def:stab}, and
$\Ebasis$ ranges over all $\Bop(\Hch)$ basis of the form
$\Ebasis =\{ \Ebe_{u} \}_{u\in\cX}$ with (\ref{eq:error_basis0})
and (\ref{eq:error_basis})
for some basis $\{ \ket{0}, \dots,\ket{\dmn-1} \}$ of $\Hch$ and 
some primitive $\dmn$-th root of unity $\omega$.
We call an $\Ebasis$-channel $\{ \sqrt{P(u)}\Ebe_{u} \}_{u\in\cX}$
satisfying 
\begin{equation} \label{eq:d-p-depo}
P(u)=\begin{cases} p/(\dmn^2-1) & \mbox{if $u\ne(0,0)$},\\
1-p & \mbox{if $u=(0,0)$}
\end{cases}
\end{equation}
for some $0 \le p \le 1$ a ($\dmn$-dimensional) $p$-depolarizing channel.

Then, we have the next.
%Then, $\Capa(\cA | \{ \rsC_n \})$ can be given as follows.

\begin{theorem}\label{prop:cond_stab}
For the $p$-depolarizing channel with $0 \le p \le 1$ %(\dmn^2-1)/\dmn^2$, ver3
%$0 \le p \le 1$ ?
and for the present choice of $\rsC_n$,
i.e., for $\rsC_n=\bigcup_{\Ebasis} \rsC_n(\Ebasis)$, we have
\begin{eqnarray*}
\Capa(\cA|\{ \rsC_n \}) 
&= &\sup_{\varnin \ge 1}
\max_{ \Hcd \in \rsC_{\varnin}}
\frac{I_{\rm c}\big((\dim \Hcd)^{-1} \Pi_{\Hcd},\cA^{\tnsr \varnin}\big)}{\varnin}\\
&= &\lim_{\varnin \to \infty}
\max_{ \Hcd \in \rsC_{\varnin}}
\frac{I_{\rm c}\big((\dim \Hcd)^{-1} \Pi_{\Hcd},\cA^{\tnsr \varnin}\big)}{\varnin}\\
&= &\lim_{\varnin \to \infty}
\max_{\Cin} \frac{\varkin-H(\overleftarrow{\Pin}|\Bar{\Pin})}{\varnin},
\end{eqnarray*}
where the maximum with respect to $\Cin$ is over all
$\Cin \subset \myF^{2\varnin}$ with $\dim \Cin=\varnin-\varkin$,
$1 \le \varkin \le \varnin$, and $\Cin \subset \Cin^{\perp}$.
Here, the probability distribution $\Pin$ is defined by (\ref{eq:Pin})
with (\ref{eq:d-p-depo}).
\enth

To prove this, 
we use the next symmetric property of the depolarizing channel:
For this channel, the representation
$\{ \sqrt{1-p} I \} \cup \{ \sqrt{p/(\dmn^2-1)} \Ebe_{u} \}_{u\in\cX\setminus\{(0,0)\}}$
does not depend on the choice of
the basis $\{ \ket{0}, \dots,\ket{\dmn-1} \}$ and $\omega$, which determine
$\Ebasis=\{\Ebe_{u}\}_{u\in\cX}$~\cite{werner01}.
Because of this property, we will obtain Theorem~\ref{prop:cond_stab}
if we show the next lemma.

\begin{lemma}\label{lem:cond_stab}
%%Let a basis $\{ \ket{0}, \dots,\ket{\dmn-1} \}$ of $\Hch$ and 
%%$\omega$ be given to determine
Let a basis
$\Ebasis =\{ \Ebe_{u} \}_{u\in\cX}$ be given through (\ref{eq:error_basis0})
and (\ref{eq:error_basis}).
For an $\Ebasis$-channel
$\cA \sim \{ \sqrt{P(u)} \Ebe_{u} \}_{u\in\cX}$,
we have
\begin{eqnarray*}
\Capa(\cA|\{ \rsC_n(\Ebasisch) \}) 
&= &\sup_{\varnin \ge 1}
\max_{ \Hcd \in \rsC_n(\Ebasisch) }
\frac{I_{\rm c}\big((\dim \Hcd)^{-1} \Pi_{\Hcd},\cA^{\tnsr \varnin}\big)}{\varnin}\\
&= &\lim_{\varnin \to \infty}
\max_{ \Hcd \in \rsC_n(\Ebasisch) }
\frac{I_{\rm c}\big((\dim \Hcd)^{-1} \Pi_{\Hcd},\cA^{\tnsr \varnin}\big)}{\varnin}\\
&= &\lim_{\varnin \to \infty}
\max_{\Cin} \frac{\varkin-H(\overleftarrow{\Pin}|\Bar{\Pin})}{\varnin},
\end{eqnarray*}
where the maximum with respect to $\Cin$ is over all
$\Cin \subset \myF^{2\varnin}$ with $\dim \Cin=\varnin-\varkin$,
$1 \le \varkin \le \varnin$, and $\Cin \subset \Cin^{\perp}$.
Here, the probability distribution $\Pin$ is given by (\ref{eq:Pin}).
\enlem

A poof of this lemma is given in Appendix~\ref{ss:proof_cond}.
From the proof,
it is clear that Theorem~\ref{prop:cond_stab} remains true
even if we extend $\rsC_n=\bigcup_{\Ebasis} \rsC_n(\Ebasis)$ so that
it includes the symplectic codes designed with
$\Bar{\Ebasis}_n = \{ \Ebe_{x_1}^{(1)} \tnsr \cdots \tnsr \Ebe_{x_n}^{(n)}
\mid (x_1,\dots,x_n) \in \cX^n \}$ instead of $\Ebasis_n$,
where each $\{ \Ebe_{u}^{(i)}\}$ is defined as $\Ebasis$ with
some basis $\{ \ket{0}, \dots, \ket{\dmn-1} \}$ and some $\omega$, which
may vary according to $i$.
%from place to place. %with the $n$ copies of $\Hch$.

\subsection{Superadditivity of Coherent Information}

The conditional capacity in Theorem~\ref{prop:cond_stab}
is the limit of $c_n/n$, where
\begin{equation}\label{eq:c_n}
c_n = \sup_{\Hcd \in \rsC_{\varnin} }
I_{\rm c}\big((\dim \Hcd)^{-1} \Pi_{\Hcd}, \cA^{\tnsr \varnin}\big),
%=\max_{\Cso} [\varkin-H(\overleftarrow{\Pin}|\Bar{\Pin})],
\quad n=1,2,\dots.
\end{equation}
A natural question is whether $\lim_n c_n/n > c_1$ or not.
Shor and Smolin numerically demonstrated that $\lim_n c_n/n > c_1$ 
%ver2. proved 
for very noisy 2-dimensional depolarizing channels,
which showed the remarkable feature of $c_n$, or its counterpart $c'_n$
that is defined by (\ref{eq:c_n}) with $\rsC_{\varnin}$
replaced by the set of all subspaces of $\Hch^{\tnsr n}$.
Note that $\lim_n c'_n/n$ is an upper bound on the unconditional
capacity $\Capa(\cA)$ by Lemma~\ref{lem:ub4cond_cap}.
For the erasure channel, 
$\lim_n c'_n/n$ is known to equal $c'_1$~\cite{barnum98e},
%$\lim_n c'_n = c'_1$
which is indeed the capacity.

Here this paper reports that superadditivity of $c_n$ has been observed 
for very noisy $3$-dimensional $p$-depolarizing channel.
Specifically, a numerical evaluation using the repetition code
$\spn \{ 1100000$, $1010000, \dots, 1000001 \}$ as an inner code, %4LAYOUT
where $1100000 \in \SINT_9^7$ denotes the vector 
$(1,0,1,0,0,0,\dots,0,0)$ $\in \field_{3}^{14}$ and so on, %4LAYOUT
has shown that for $0.2552 \le p \le 0.2557$, $c_7>0$ while $c_1<0$. %ver3

\section{Bounds for General Discrete Channels \label{ss:gen}}

We remark that this work's bound holds true
for general discrete memoryless channels (TPCP maps) 
as treated in \cite{hamada01g}.
Namely, if we associate the probability distribution
$P=P_{\cA}$ with a channel $\cA$
[or $P=P_{\cU\cA}$ with some TPCP map $\cU$ on $\Bop(\Hch)$]
as in \cite[Section~II]{hamada01g},
then the bound in (\ref{eq:exp_1st}) and that in Corollary~\ref{coro:1}
are true for this channel. This can be shown
in a quite similar way to that in \cite{hamada01g}.
That is, if the minimum fidelity $F$ in (\ref{eq:Fbar0}) is
replaced by the minimum average fidelity $\Fav$ introduced in \cite{hamada01g},
%and evaluated instead of minimum fidelity $F$,
the same bound holds on $\Fav$ for general memoryless channels.
Then, owing to the fact \cite{hamada01g} that a lower bound on $\Fav$ 
gives asymptotically the same bound on $F$,
we obtain the lower bound on the minimum fidelity $F$ of the best codes
used on general channels.
These bounds also apply to
`blockwise' memoryless channels (TPCP maps)
$\cA_n$ on $\Bop(\Hch^{\tnsr n})$ if we associate
the probability distribution $P_n=P_{\cA_n}$ on $\cX^n$
[or $P_n=P_{\cU\cA_n}$ with some TPCP map $\cU$ on $\Bop(\Hch^{\tnsr n})$]
with $\cA_n$ as in \cite[Section~V, Definition~2]{hamada01g}
and use $P_n$ in place of $P^n$ in (\ref{eq:Pin}).

\section{Concluding Remarks \label{ss:conc}}

This paper has presented a lower bound on the quantum capacity
which has a close relation to the known upper bound based on coherent
information~\cite{barnum00}.
%As already mentioned, 
This author conjectures that this bound is actually the conditional capacity
of general Pauli or $\Ebasis$-channels on all symplectic stabilizer codes.
It might even be true that the lower bound is tight 
as one on the usual (unconditional) quantum capacity for Pauli channels,
which would be proved by showing that the maximum of coherent information
were nearly %(or exactly) 
achieved by an input state proportional to the projections onto
the code space of a symplectic stabilizer code for large enough $n$.
%infinity many values of $n$.
%
If the quantum capacity were proven to be the 
coherent-information upper bound $\lim_n c'_n/n$,
it would still leave room for investigation
since the bound is a limiting expression and
we do not know how to calculate it except for few cases~\cite{barnum98e}.

In the previous work~\cite{hamada01e}, this author conjectured
that the exponent appearing in the fidelity bound in \cite{hamada01e}
is not the optimum for some channels. 
Now this fact has been established at least numerically 
%in view of (\ref{eq:exp_1st})
since the bound in \cite{hamada01e}
is the same as the right-hand side of (\ref{eq:exp_1st})
%viz., $\Enk{n}{k}(R,P)$ 
when $n=k=1$, in which case $P_{\Cso}=P$, 
and we have Shor and Smolin's
numerical evaluation for the depolarizing channel, %in (\ref{eq:depo}),
from which it follows that there exist some $\varnin \ge 3$ and
relatively large $p$ such that
$\Enk{\varnin}{1}(R\varnin,\Pin)$ is positive while $\Enk{1}{1}(R,P)$
vanishes. The problem of determining the quantity 
in the left-hand side of (\ref{eq:exp_1st}) would deserve investigations
in view of the great attention paid to the
corresponding problem in classical information theory; 
an improvement on $\Enk{1}{1}(R,P)$
%for channels of low noise-level 
for $P$ with large $P\big((0,0)\big)$ can be found in \cite{barg02}.

\section*{Acknowledgment}

The author wishes to thank
Hiroshi Imai and Keiji Matsumoto of the QCI project for support. 

\appendices

\mysectionapp{Proof of Lemma~\ref{lem:hyperbolic_plane} \label{ss:proof4hp}}

In this proof, we use the property of a nondegenerate
bilinear form $(\cdot,\cdot)'$ on $V$ with $\dim V=2l$ that
\begin{equation}\label{eq:property_blform}
\dim W + \dim W^{\perp'} = 2l
\end{equation}
for any subspace $W\subset V$, where
$W^{\perp'} = \{ y\in V \mid \forall x\in W,(x,y)'=0 \}$~\cite{artin,grove}.
We construct pairs $(g_i,h_i)$ of the property in (\ref{eq:hyperbolic_plane})
%or Lemma~ %but in this tex style, equations are easier to find 
in the following two-stage algorithmic process.
%
%\noindent 
\begin{list}{}{\setlength{\leftmargin}{8mm}}
\item[(i)] Put $\varl=n-k$, $V=\myF^{2n}$ and $W=\Cso$. Repeat Procedure~1.\\
{\em Procedure~1.}\/
If $\varl=0$, then %stop this procedure and 
go to (ii).
Since $W$ is contained in 
$W^{\perp} \cap V$ and 
$0< \dim W^{\perp} \cap V=2k+\varl <\dim V=2(k+\varl)$ by (\ref{eq:property_blform}), where in this case the bilinear form $(\cdot,\cdot)'$ is 
simply the restriction of $\syp{\cdot}{\cdot}$ to $V$ (see also \cite[Proposition~2.9]{grove}),
there is a vector $h_{\varl} \in V \setminus W^{\perp}$ with
$\syp{g_{\varl}}{h_{\varl}}=1$.
Define a subspace $V'\subset\myF^{2n}$ by $V=
V' \perp \spn\{ g_{\varl},h_{\varl} \}$, replace $V$ with $V'$,
and replace $W$ with $\spn \{ g_1,\dots,g_{\varl-1}\}$.
Decrease $\varl$ by 1.
\end{list}
%
%We repeat Procedure~1 until $\varl$ becomes $0$. %$n-k$ times. 
Up to now, we have hyperbolic pairs $(g_1,h_1), \dots, (g_{n-k},h_{n-k})$.

\noindent
\begin{list}{}{\setlength{\leftmargin}{8mm}}
\item[(ii)] Put $m=n-k$. Repeat Procedure~2.\\
{\em Procedure~2.}\/
If $m=n$, terminate this procedure.
Let $V$ be defined by $\myF^{2n} = V \perp \spn \{
g_1,h_1,\dots,$ $g_{m},h_{m}\}$. %4LAYOUT
Choose an arbitrary nonzero vector $g_{m+1}\in V$.
Since $W=\spn \{ g_{m+1} \}$ is contained in 
$W^{\perp}\cap V$ and $\dim W^{\perp} \cap V =2(n-m)-1< \dim V=2(n-m)$,
there is a vector $h_{m+1} \in V \setminus W^{\perp}$ with
$\syp{g_{m+1}}{h_{m+1}}=1$. 
Increase $m$ by 1.
\end{list}
%
%$We repeat Procedure~2 until $m$ becomes $0$.
%
Thus, we have hyperbolic pairs $(g_1,h_1), \dots, (g_{n},h_{n})$.

\mysectionapp{Proof of Lemma~\ref{lem:te} \label{ss:proof4te}}

In this proof,
when $x\in\myF^{2n}$ is of the form $x=\sum_{i=1}^{n} z_i h_i$,
we write
$
\Ebesp{x}=\prod_{i=1}^{n} (\Ebe_{h_i})^{z_i} %ver2
$
where the product on the left-hand side is unambiguous because
$(\Ebe_{h_i})^{z_i}$, $i=1,\dots,n$, commute with each other.
Similarly, when $x\in\myF^{2n}$ is of the form $x=\sum_{i=1}^{n} w_i g_i$,
we write
$
\Ebesp{x}=\prod_{i=1}^{n} (\Ebe_{g_i})^{w_i}. %ver2
$
Due to (\ref{eq:propertyEB}), $\Ebe_{x}$
and $M=\Ebesp{\smlsum_i u_i h_{i+n-k}} \Ebesp{\smlsum_i u'_i
g_{i+n-k}} \Ebeh{\vartpr} \Ebeg{t}$, 
where $i$ runs from $1$ to $k$ in the summations, are the same up to a
factor of modulus one that solely depends on $x$, so that 
$\Ebe_{x}\rho\Ebe_{x}^{\dagger}=M\rho M^{\dagger}$.
Hence, if $\Bar{X}_{u}$ and $\Bar{Z}_{u'}$ differ only by phase factors
from $\Ebesp{\smlsum_i u_i h_{i+n-k}}$ and $\Ebesp{\smlsum_i u'_i
g_{i+n-k}}$, respectively, 
and the factors do not depend on $(b_i)_{1\le i \le k}$
when they act on $\ket{\ghb{z_1,\dots,z_{n-k},b_1,\dots,b_k}}$,
then we will obtain the lemma.
%since an operator $M$ acts as $M \rho M^{\dagger}$
%on a density operator $\rho$. 

%The action of $\Bar{X}_{u} \Bar{Z}_{u'}$ can be understood as follows.
It is seen from %(\ref{eq:only_h}) and 
(\ref{eq:basis4Hn}) 
that the action of $\Bar{X}_{u}$ is the same as 
$\Ebesp{\smlsum_{i} u_i h_{i+n-k}}$.
On the other hand, the action of $\Bar{Z}_{u'}$ is the same as %that of
$\Ebesp{\smlsum_{i} u'_i g_{i+n-k}}$
up to an irrelevant phase factor.
This can be seen by
the following chain of equalities, where (\ref{eq:commutative}),
(\ref{eq:eigenspace}) and (\ref{eq:basis4Hn}) are used, %{eq:codespace2}?
and in all summations,
$i$ runs from $1$ to $k$, and $j$ from $1$ to $n-k$,
and $\lambda,\lambda'$, which depend on $x$ and $b_i$, are defined by
\[
\Ebesp{\smlsum_{j} z_j h_j + \smlsum_i b_i h_{i+n-k}}
= \lambda 
\Ebe_{\smlsum_{j} z_j h_j + \smlsum_i b_i h_{i+n-k}}
\]
and
\[
\Ebesp{\smlsum_i u'_i g_{i+n-k}}
= \lambda' 
\Ebe_{\smlsum_i u'_i g_{i+n-k}},
\]
which is possible owing to (\ref{eq:propertyEB}):
\begin{eqnarray*}
\lefteqn{\Ebesp{\smlsum_i u'_i g_{i+n-k}} \ket{\ghb{z_1,\dots,z_{n-k},b_1,\dots,b_k}}}&&\\
&=&\Ebesp{\smlsum_i u'_i g_{i+n-k}} \Ebesp{\smlsum_{j} z_j h_j + \smlsum_i b_i h_{i+n-k}}
\ket{\ghb{0,\dots,0}}\\
%&=&\prod_{i} \Ebe_{g_{i+n-k}}^{u'_i} \prod_{j} \Ebe_{h_j}^{z_j}  \prod_i \Ebe_{h_{i+n-k}}^{b_i} \ket{\ghb{0,\dots,0}}\\
&=& \lambda \lambda'
\Ebe_{\smlsum_i u'_i g_{i+n-k}} \Ebe_{\smlsum_{j} z_j h_j + \smlsum_i b_i h_{i+n-k}} 
\ket{\ghb{0,\dots,0}}\\
&=& \lambda \lambda' \omega^{\syp{\smlsum_i u'_i g_{i+n-k}}{\,\smlsum_{j} z_j h_j + \smlsum_i b_i h_{i+n-k}}}
\Ebe_{\smlsum_{j} z_j h_j + \smlsum_i b_i h_{i+n-k}} \Ebe_{\smlsum_i u'_i g_{i+n-k}} 
\ket{\ghb{0,\dots,0}}\\
&=& \omega^{\syp{\smlsum_i u'_i g_{i+n-k}}{\,\smlsum_{j} z_j h_j + \smlsum_i b_i h_{i+n-k}}}
\Ebesp{\smlsum_{j} z_j h_j + \smlsum_i b_i h_{i+n-k}} \Ebesp{\smlsum_i u'_i g_{i+n-k}}
\ket{\ghb{0,\dots,0}}\\
&=& \omega^{\smlsum_i u'_ib_i}
\Ebesp{\smlsum_{j} z_j h_j + \smlsum_i b_i h_{i+n-k}} \Ebesp{\smlsum_i u'_i g_{i+n-k}}
\ket{\ghb{0,\dots,0}}\\
&=& \omega^{\smlsum_i u'_ib_i}
\Ebesp{\smlsum_{j} z_j h_j + \smlsum_i b_i h_{i+n-k}} 
\prod_i (\Ebe_{g_{i+n-k}})^{u'_i}
\ket{\ghb{0,\dots,0}}\\
&=& \omega^{\smlsum_i u'_ib_i+ \smlsum_i u'_i \mu_{i+n-k}} 
\Ebesp{\smlsum_{j} z_j h_j + \smlsum_i b_i h_{i+n-k}}
\ket{\ghb{0,\dots,0}}\\
&=& \omega^{\smlsum_i u'_ib_i + \smlsum_i u'_i \mu_{i+n-k}}
\ket{\ghb{z_1,\dots,z_{n-k},b_1,\dots,b_k}}.
\end{eqnarray*}

\mysectionapp{Proof of Theorem~\ref{th:main2} \label{ss:proof2}}

In this appendix, the lower bound in Theorem~\ref{th:main2} will
be established using
the concatenated code in Section~\ref{ss:cat} as well as the notation
therein.
In the proof, 
the random coding proof method is employed, where as in DiVincenzo, Shor and
Smolin~\cite{ss97,dss98},
{\em an inner code $\Cso$ having parameters $\varkin$ and $\varnin$, %the ver3
$1\le\varkin\le\varnin$, is fixed}\/ %arbitrarily fixed %ver3 
and the average fidelity over all possible outer codes $\Cout$
is evaluated. Namely, 
we evaluate
\begin{equation}\label{eq:Fbar0}
\Fbar = \frac{1}{\crd{\Aso}} \sum_{\Cout\in\Aso} F\big(\Hcd(\Cout),
\cR\cA^{\tnsr \varnin\varN}\big),
\end{equation}
where $F$ is defined in (\ref{eq:defF}), 
the ensemble $\Aso$ is specified below in (\ref{eq:Aso}),
and $\Hcd(\Cout)$ is one of
the $\dmn^{n\varN-\varK}$ symplectic quantum codes
of dimension $\dmn^{\varK}$ associated with $\cnc{\Cin}{\Cout}$,
and $\cR$ is determined from $\Icr_0$ which will be given below.
It will be helpful to notice
that $\spn\Gin$ is contained in $\cnc{\Cin}{\Cout}$,
which, in turn, is contained in $\cnc{\Cin}{\Cout}^{\perp}$,
so that any coset of $\cnc{\Cin}{\Cout}^{\perp}$ in
$\myF^{2\varnin \varN}$ is a union of some cosets of $\spn\Gin$.
We will work largely with cosets of $\spn\Gin$ rather than individual
sequences in $\myF^{2\varN\varnin}$ because due to Lemma~\ref{lem:te},
error operators indexed by sequences in a fixed coset act on the states
exactly in the same way.

It is convenient to view
\begin{equation*} %\label{eq:x0}
y =
(y_{1,1},\dots,y_{1,2\varnin},\dots, y_{\varN,1},\dots,y_{\varN,2\varnin})
\in \myF^{2\varnin \varN}
\end{equation*}
in terms of the  basis 
\[
\bigcup_{j=1}^{\varN} \bigcup_{i=1}^{\varnin} \{ g_i^{(j)}, h_i^{(j)} \}.
\]
Let us expand $y \in \myF^{2\varnin}$ as
\begin{equation}\label{eq:x1}
y = \sum_{j=1}^{\varN} \sum_{i=1}^{\varnin} w_{j,i} g_i^{(j)} + z_{j,i} h_i^{(j)}
\end{equation}
and consider the transformation that maps $y$ to
\begin{equation} \label{eq:x3}
y'= (\tvarz_{1,1},\tvarw_{1,1},\dots,\tvarz_{1,\varnin},\tvarw_{1,\varnin},\dots,
\tvarz_{\varN,1},\tvarw_{\varN,1},\dots,\tvarz_{\varN,\varnin},\tvarw_{\varN,\varnin}).
\end{equation}
Then, it is easy to see that each blocks of length $2n$ suffers the transformation $\coch$ defined by (\ref{eq:coch}):
\begin{equation*} %\label{eq:x2}
(\tvarz_{j,1},\tvarw_{j,1},\dots,\tvarz_{j,\varnin},\tvarw_{j,\varnin})=\coch\big((y_{j,1},\dots, y_{j,2\varnin})\big),
\quad j=1,\dots,\varN.
\end{equation*}

Note that the vectors $y\in\myF^{2\varnin \varN}$
which, when expanded as in (\ref{eq:x1}),  
have the same $z_{j,i}$ for $j=1,\dots,\varN$, $i=1,\dots,\varnin$,
and the same $w_{j,i}$ for $j=1,\dots,\varN$, $i=\varnin-\varkin+1,\dots,\varnin$,
form a coset of $\spGin$ in $\myF^{2\varnin \varN}$. 
In other words, the set
\[
\{ y \mid 
\syp{g_i^{(j)}}{y} = \tilde{z}_{j,i}, \, 1 \le j \le \varN,\,
1 \le i \le \varnin; \, 
\syp{y}{h_i^{(j)}} = \tilde{w}_{j,i}, \, 1\le j \le \varN,\,
\varnin-\varkin+1 \le i \le \varnin  \}
\]
for a fixed pair
$\big((\tilde{z}_{j,i}), (\tilde{w}_{j,i})\big) \in (\myF^{n})^{\varN}
\times (\myF^{k})^{\varN}$ is a coset of $\spGin$.
With the decomposition of an error operator in Lemma~\ref{lem:te} in mind,
we rather write a coset of $\spGin$ as
\[
\{ y \mid z(y)=\tilde{z},\, v(y) = \tilde{\varv} \}
\]
with a fixed pair
$(\tilde{z},\tilde{\varv}) \in (\myF^{n-k})^{\varN}
\times (\myF^{2k})^{\varN}$,
where the two sequences $z(y)$ and $\varv(y)$ are defined by
\begin{equation}\label{eq:y0}
z(y)=(z_1,\dots,z_{\varN})\in(\myF^{\varnin-\varkin})^{\varN} \quad \mbox{and}\quad
\varv(y)=(\varv_1,\dots,\varv_{\varN})\in(\myF^{2\varkin})^{\varN}, 
\end{equation}
and
\begin{equation}\label{eq:y1}
z_j=(z_{j,1},\dots,z_{j,\varnin-\varkin})\quad \mbox{and}\quad
\varv_j=(\tvarz_{j,\varnin-\varkin+1},\tvarw_{j,\varnin -\varkin+1},
\dots,\tvarz_{j,\varnin},\tvarw_{j,\varnin}), \quad j=1,\dots,\varN,
\end{equation}
with (\ref{eq:x1}).
We denote 
by $\cset{\Gin}{z(y),\varv(y)}$ the %corresponding
coset of $\spGin$ in $\myF^{2\varnin \varN}$ that contains $y$
for any $y\in\myF^{2\varnin\varN}$.
The simplest coset representative of
a coset $\cset{}{\tilde{z},\tilde{v}}$ is
%where $(\tilde{z},\tilde{\varv}) \in (\myF^{n-k})^{\varN}
%\times (\myF^{2k})^{\varN}$,
the vector $y$ with
$w_{j,i}=0$ for $1 \le j \le \varN$, $1 \le i \le \varnin-\varkin$
when represented as in (\ref{eq:x1}).
The set (transversal) consisting of these coset representatives is denoted by $\yg$.
On the other hand, 
cosets of $\cnc{\Cin}{\Cout}^{\perp}$ in $\myF^{2\varnin\varN}$
can be specified as follows in view of (\ref{eq:coset}):
A coset of $\cnc{\Cin}{\Cout}^{\perp}$ has the form
\begin{gather}
\{ y \mid \syp{g_i^{(j)}}{y} = \tilde{z}_{j,i}, \, 1 \le j \le \varN,\,
1 \le i \le \varnin -\varkin;\,
\syp{\Bar{g'_i}}{y} = \varz_{i},\, 1 \le i \le \varkin \varN -\varK \}\notag\\
= 
\{ y \mid z(y) = \tilde{z};\,
\syp{\Bar{g'_i}}{y} = \varz_{i},\, 1 \le i \le \varkin \varN -\varK \}\label{eq:large_coset}
\end{gather}
with fixed 
$\tilde{z}=\big((\tilde{z}_{j,i})_{1\le i\le n-k}\big)_{1\le j \le \varN}$
and $\varz=(\varz_i)_{1 \le i \le k \varN -\varK}$. We denote this coset
by $\Lcset{}{\tilde{z},\varz}$.

Given $z=(z_1,\dots,z_{\varN})\in(\myF^{\varnin-\varkin})^{\varN}$ and
$\varv=(\varv_1,\dots,\varv_{\varN})\in(\myF^{2\varkin})^{\varN}$,
we denote the rearranged sequence
$\big((z_1,\varv_1),\dots,(z_{\varN},\varv_{\varN})\big)\in  (\myF^{\varnin-\varkin} \times
\myF^{2\varkin})^{\varN}$ by $[z,\varv]$,
and define a probability distribution $\sP_{z,\varv}$,
which is called the {\em type}\/ of the sequence $[z,v]$, by
\begin{equation}\label{eq:type}
\sP_{z,\varv}(s,\varu)= \frac{\crd{ \{ i \mid  (z_i,\varv_i)= (s,\varu),\,  1 \le i
\le \varN \}}}{\varN},\quad
s \in\myF^{\varnin-\varkin},\, \varu\in\myF^{2\varkin},
\end{equation}
and put
\begin{gather*}
\sP_{z}(s) = \sum_{\varu\in \myF^{2\varkin}} \sP_{z,\varv}(s,\varu),\quad s \in\myF^{\varnin-\varkin},\\
\sP_{v}(\varu) = \sum_{s\in\myF^{\varnin-\varkin}} \sP_{z,\varv}(s,\varu),\quad \varu\in\myF^{2\varkin},
\end{gather*}
which are the types of $z$ and $v$, respectively.

To make use of Lemma~\ref{lem:coset_leaders},
we choose a representative from each coset $\Lcset{}{\zpr,\varzpr}$ as follows.
Among those sequences $y$ that belong to 
both $\yg$ and $\Lcset{}{\zpr,\varzpr}$,
we choose one that minimizes 
$H_{\rm c}(\sP_{\zpr,\varv(y)})=H(\overleftarrow{\sP_{\zpr,\varv(y)}}|\sP_{\zpr})$,
where $H_{\rm c}(Q)$ is shorthand for
$H(\overleftarrow{Q}|\Bar{Q})$.
We apply Lemma~\ref{lem:coset_leaders}
defining $\Icr_0=\Icr_0(\Cout)$ as the set of these representatives.
Denote $\Icr$ in the lemma by $\Icr(\Cout)$, viz.,
\begin{equation}\label{eq:Icr_cat}
\Icr(\Cout) = \Icr_0(\Cout)+\cnc{\Cin}{\Cout},
\end{equation}
and put
\begin{equation}\label{eq:Aso}
\Aso  = \{ \Cout \subset \myF^{2 \varkin\varN} \mid \mbox{$\Cout$ linear}, \ \Cout \subset
\Cout^{\perp},\ \dim \Cout = \varkin\varN-\varK \}.
%\mbox{where $m=\varnin \varN$}.
\end{equation}
%and put $\Aso=\Aso_{\varkin \varN, k}$.
Then, we have
\begin{eqnarray}
1- \Fbar & \le & \frac{1}{\crd{\Aso}} \sum_{\Cout\in\Aso} \sum_{y
     \notin \Icr(\Cout)}  {P}^{\varnin \varN}(y) \nonumber\\
     & = & \frac{1}{\crd{\Aso}} \sum_{\Cout\in\Aso} \sum_{y \in
     \myF^{2\varnin \varN}  }
    {P}^{\varnin \varN}(y) \indc [ y \notin \Icr(\Cout) ]  \nonumber\\
%     & = & \frac{1}{\crd{\Aso}} \sum_{y \in \myF^{2n}} P^{\varN}(y) \sum_{\Cout\in\Aso} \indc [ y \notin \Icr(\Cout) ] \nonumber\\ %4SPACE
     & = & \sum_{y \in \myF^{2\varnin \varN}} {P}^{\varnin \varN}(y) \frac{\crd{\Bcn(y)}}{\crd{\Aso}},
\label{eq:pr0}
\end{eqnarray}
where $\indc[T]=1$ if a statement $T$ is true and $\indc[T]=0$ otherwise,
and
\[
\Bcn(y) = \{ \Cout \in \Aso \mid y \notin \Icr(\Cout) \}, \quad
y\in\myF^{2\varnin \varN}.
\]
The fraction $\crd{\Bcn(y)}/\crd{\Aso}$ is trivially bounded as
\begin{equation} \label{eq:pr1}
\frac{\crd{\Bcn(y)}}{\crd{\Aso}} \le 1, \quad y \in\myF^{2\varnin\varN}.
\end{equation}
%Putting $m=\varnin \varN$, 
We use the next lemma~\cite{hamada01g},
which is a variant of a fact established by Calderbank {\em et al.}~\cite{crss97}.

\begin{lemma}\label{lem:C_uniform}
Let
\[
\Acn(x) = \big\{ \Cout \in \Aso \mid x \in \Cout^{\perp} \setminus \{ 0 \} \big\}.
\]
Then, $\crd{\Acn(0)}=0$ and
\begin{equation*} %\label{eq:C_uniform}
\frac{\crd{\Acn(x)}}{\crd{\Aso}}
 = \frac{\dmn^{\varkin\varN+\varK} - 1}{\dmn^{2\varkin\varN}-1}  
\le \frac{1}{\dmn^{\varkin\varN-\varK}}, \quad x\in\myF^{2\varkin\varN},\
 x \ne 0.
\end{equation*}
\enlem

From the design of $\Icr_0(\Cout)$
specified above,
it follows that
\begin{equation}\label{eq:Bcnbound}
\Bcn(y) \subset \big\{ \Cout \in \Aso \mid \exists \varv'
\in\myF^{2\varkin \varN},\, H_{\rm c}(\sP_{z(y),\varv'}) \le H_{\rm
c}(\sP_{z(y),\varv(y)}),\, \varv'-\varv(y)\in \Cout^{\perp} \setminus \{
0 \} \big\}.
\end{equation}
[This can be seen as follows.
First, we consider the case where $y\in\yg$.
Recall $\Lcset{}{\tilde{z},\varz}$ was defined by
(\ref{eq:large_coset}), and observe
$\syp{g'_i}{v(y)}=\syp{\Bar{g'_i}}{\Bar{v(y)}}=\syp{\Bar{g'_i}}{y}$
from (\ref{eq:bar_g}),
where again $v=v(y)$ and $z=z(y)$ are specified by
(\ref{eq:x1}), (\ref{eq:y0}) and (\ref{eq:y1}).
Hence, $x\in\yg$ and $y\in\yg$ are in the same coset of
$\cnc{\Cin}{\Cout}^{\perp}$ if and only if 
\[
z(x)=z(y)\quad \mbox{and}\quad
\syp{g'_i}{v(x)}=\syp{g'_i}{v(y)},\
1 \le i \le \varkin \varN -\varK
\]
by (\ref{eq:large_coset}),
which can be restated as
\[
z(x)=z(y) \quad
\mbox{and}\quad \varv(x)-\varv(y)\in \Cout^{\perp}.
\]
Since, at the beginning of the paragraph containing (\ref{eq:Icr_cat}),
we have chosen a coset representative
%$x\in\Icr_0(\Cout)$ 
that minimizes $H_{\rm c}$ in $\Lcset{}{\tilde{z},\varz}\cap\yg$
for each coset $\Lcset{}{\tilde{z},\varz}$,
it follows that for any $y\in\Lcset{}{z(y),\varz}\cap\yg$,
the condition $y \notin \Icr(\Cout)$ occurs only if there exists a vector
other than $y$ in $\Lcset{}{z(y),\varz}\cap\yg$
that has conditional entropy $H_{\rm c}$ as small as $y$,
which implies (\ref{eq:Bcnbound}).
To see this for a general $y\in\myF^{2n\varN}$, note that
%$\Cout\in\Bcn(y)$ if and only if $\Cout\in\Bcn(x)$ for
%some (hence, all) $x$ with $x-y\in\cnc{\Cin}{\Cout}$. In other words, 
for a fixed coset of $\cnc{\Cin}{\Cout}$,
either each element $y$ of the coset satisfies $\Cout\in\Bcn(y)$ or
each satisfies $\Cout\notin\Bcn(y)$ because
of Lemma~\ref{lem:coset_leaders},
or specifically, (\ref{eq:Icr_cat}) in this case.
Especially, $\Cout\in\Bcn(y)$ if and only if $\Cout\in\Bcn(x)$
for the vector $\hat{y}\in\yg$ with $\hat{y}-y\in\spGin$,
so that we can judge whether $y \notin \Icr(\Cout)$ occurs
or not by checking the condition $\hat{y} \notin \Icr(\Cout)$
for $\hat{y}\in\yg$ with $\hat{y}-y\in\spGin$.]

Owing to (\ref{eq:Bcnbound}), we have 
\begin{eqnarray}
 \crd{\Bcn(y)} &\le &\sum_{\varv'\in (\myF^{2\varkin})^{\varN} :\, H_{\rm c}(\sP_{z(y),\varv'}) \le  H_{\rm c}(\sP_{z(y),\varv(y)})}
   \crd{\Acn\big(v'-v(y)\big)}\nonumber\\
  &\le & \sum_{\varv'\in (\myF^{2\varkin})^{\varN} :\, H_{\rm c}(\sP_{z(y),\varv'}) \le  H_{\rm c}(\sP_{z(y),\varv(y)})}
\crd{\Aso}{\dmn}^{-\varkin\varN  + \varK}, \label{eq:pr2} 
\end{eqnarray}
where the second inequality is due to Lemma~\ref{lem:C_uniform}.
Then, from (\ref{eq:pr0}), (\ref{eq:pr1}) and (\ref{eq:pr2}), it
follows that
\begin{equation*}
1-\Fbar \le
\sum_{z\in \myF^{(\varnin-\varkin)\varN} }
\sum_{v\in \myF^{2\varkin \varN}}
\sum_{y \in \cset{}{z,v}} P^{\varnin \varN} (y) \ \min \Biggl\{ \
 \sum_{\varv'\in\myF^{2\varkin \varN} :\, H_{\rm c}(\sP_{z,\varv'})
\le  H_{\rm c}(\sP_{z,\varv})} \dmn^{-(\varkin\varN-\varK)},\ 1 \
\Biggr\}.
\end{equation*}
Recalling the probability distribution $P_{\Cso}$ defined by (\ref{eq:Pin})
and the transformation that converts $y$ into $y'$ in (\ref{eq:x3}),
we have
\[
\Pin^{\varN} ([z,\varv])= \sum_{y:\, y \in \cset{\Gin}{z,\varv} } P^{
\varnin \varN}(y), \quad z\in(\myF^{n-k})^{\varN}, v\in(\myF^{2k})^{\varN},
\]
where $\Pin^{\varN}$ denotes the product of $\varN$ copies of $\Pin$,
and hence, the above bound can be rewritten as
\begin{equation}\label{eq:Fbar1}
1-\Fbar \le
\sum_{z\in \myF^{(\varnin-\varkin)\varN} }
\sum_{v\in \myF^{2\varkin \varN}}
\Pin^{\varN} ([z,\varv])
 \min \Biggl\{ \
 \sum_{\varv'\in\myF^{2\varkin \varN} :\, H_{\rm c}(\sP_{z,\varv'})
\le  H_{\rm c}(\sP_{z,\varv})} \dmn^{-(\varkin\varN-\varK)},\ 1 \
\Biggr\}.
\end{equation}

Now, we will go into an argument using the method of types~\cite{csiszar_koerner,csiszar98}.
We put
\[
\tcQ_{\varN}= \tcQ_{\varN}(\myF^{n-k})= 
\{ \sP_z \mid z \in(\myF^{n-k})^{\varN} \}.
\]
For a type $Q\in\tcQ_{\varN}$, we define a set of stochastic matrices
$\cndV_{\varN}(Q)$ by
\[
\cndV_{\varN}(Q) = \{ \varW \mid 
            \exists z \in(\myF^{n-k})^{\varN}, \exists
             v \in(\myF^{2k})^{\varN},\ \sP_{z}=Q \,
            \ {\rm and} \ \,
            \overleftarrow{\sP_{z,v}} = \varW \},
\]
and put
\[
\cardcndV = \max_{Q\in\tcQ_{\varN}} |\cndV_{\varN}(Q)|.
\]
Here,
%we $\supp Q$ denotes $\{ z \in \myF^{n-k} \mid Q(z)>0 \}$ as usual,
the probability distribution $\varW(\cdot|s)$ is allowed to be undefined for
some (but not all) $s\in\myF^{n-k}$,
%and $\eff V$ denote the set of $s$ for which $V(\cdot|s)$ is defined;
the equality between stochastic
matrices $\varW$ and $\varW'$ means
$\varW(u|s)=\varW'(u|s)$ for all $u,s$ for which 
either $\varW(u|s)$ or $\varW'(u|s)$ is defined.
For a type $Q\in\tcQ_{\varN}$ of a sequence in $(\myF^{n-k})^{\varN}$
and $\varW\in\cndV_{\varN}(Q)$,
a probability distribution $Q\pdsm \varW$ on $\myF^{n-k}\times\myF^{2k}$
is defined by
\[
[Q\pdsm \varW]\big((s,u)\big)=Q(s)\varW(u|s), \quad s\in \myF^{n-k}, u\in \myF^{2k},
\]
which is the type of a sequence in $(\myF^{n-k}\times\myF^{2k})^{\varN}$.
Here we understand $[Q\pdsm \varW]\big((s,u)\big)=0$ for $s$ with $Q(s)=0$.
The set of all possible types $Q\pdsm\varW$ of sequences
$[z,v]\in(\myF^{n-k}\times\myF^{2k})^{\varN}$ is denoted by
\[
\tcQ_{\varN}(\myF^{n-k}\times\myF^{2k}).
\]
For $Q\in\tcQ_{\varN}$ and $\varW\in\cndV_{\varN}(Q)$, i.e.,
for $Q\pdsm\varW \in \tcQ_{\varN}(\myF^{n-k}\times\myF^{2k})$,
a set of sequences $\cT^{\varN}_{Q\pdsm \varW}$ is defined by
\[
\cT^{\varN}_{Q\pdsm \varW}=\{ [z,v] \in (\myF^{n-k}\times\myF^{2k})^{\varN} \mid \sP_{z,v}= Q\pdsm \varW \}.
\]
Hereafter, we write $\Pin(s,v)$ and $[Q\times \varW](s,v)$
in place of $\Pin\big((s,v)\big)$ and $[Q\times \varW]\big((s,v)\big)$, respectively.
For a fixed sequence $z\in\myF^{n-k}$, and a stochastic matrix
$V\in\cndV_{\varN}(\sP_z)$, we define
\[
\cT^{\varN}_{V}(z)
=\{ v \in (\myF^{2k})^{\varN} \mid \overleftarrow{\sP_{z,v}} =  \varW \},
%& = & \{ v \in (\myF^{2k})^{\varN} \mid \sP_{z,v} = \sP_{z}\pdsm \varW \}\\
%& = & \{ v \in (\myF^{2k})^{\varN} \mid \overleftarrow{\sP_{z,v}} =  \varW \},
\]
which is called the $\varW$-shell of $z$~\cite{csiszar_koerner}.
Clearly, 
the cardinality of $\cT^{\varN}_{\varW}(z)$ is uniform over sequences $z$ %for?
of a fixed type $Q$, and hence, we can put
\[
T^{\varN}_{V}(Q)=\crd{\cT^{\varN}_{V}(z)},
\]
where $\sP_{z}=Q$.
We use the following two basic estimates~\cite[Lemmas~2.5 and 2.6]{csiszar_koerner}, \cite[Eqs.~(II.5) and (II.7)]{csiszar98}:
\begin{eqnarray}
\Prob \{ \sP_{\rvZ,\rvV} = Q\pdsm\varW \}
&=&\crd{\cT_{Q\pdsm \varW}^{\varN}} 
\prod_{(s,\varu)\in(\myF^{\varnin-\varkin}\times \myF^{2\varkin})}
\Pin(s,\varu)^{\varN[Q\pdsm \varW](s,\varu)} \nonumber\\
&\le & \exp_{\dmn} [ -\varN D(Q\pdsm \varW || \Pin) ], \label{eq:type1}
\end{eqnarray}
where $Q\in \tcQ_{\varN}, \varW \in \cndV_{\varN}(Q)$
and the sequence of random variables $[\rvZ,\rvV]$ that takes values
in $(\myF^{\varnin-\varkin}\times \myF^{2\varkin})^{\varN}$
is drawn according to $\Pin^{\varN}$;
\begin{equation}
\cardVsh{Q}{\varW} \le \exp_{\dmn} [\varN H(\varW|Q)], \quad
Q\in \tcQ_{\varN}, \varW \in \cndV_{\varN}(Q). \label{eq:type2}
\end{equation}

We arbitrarily fix $R$, $0\le R <1$, and put
\[
\varK=\lceil kR\varN \rceil,
\]
so that the information rate $\varK/(\varnin\varN)$ of the concatenated
code is not less than $kR/\varnin$.
From (\ref{eq:Fbar1}), (\ref{eq:type1}), (\ref{eq:type2}) and the inequality
$\min \{ a+b, 1\} \le \min \{ a, 1\} + \min \{ b, 1\}$ for $a,b \ge 0$,
we have
\begin{eqnarray*}
1-\Fbar &\le& \sum_{[z,\varv]\in(\myF^{\varnin-\varkin}\times \myF^{2\varkin})^{\varN}} \Pin^{\varN}([z,\varv]) \ \min \Biggl\{ \
 \sum_{\varv'\in\myF^{2\varkin \varN} :\, H_{\rm c}(\sP_{z,\varv'}) \le  H_{\rm c}(\sP_{z,\varv})} \dmn^{-(\varN\varkin-\varK)},\ 1 \ \Biggr\}\\
 & \le & \sum_{Q\in \tcQ_{\varN}} \sum_{\varW \in \cndV_{\varN}(Q)} \crd{\cT_{Q\pdsm
 \varW}^{\varN}} \prod_{(s,\varu)\in (\myF^{\varnin-\varkin}\times \myF^{2\varkin})} \Pin(s,\varu)^{\varN[Q\pdsm \varW](s,\varu)}\\
&&\,\,\,\,\, \,\,\,\,\, \,\,\,\,\, \,\,\,\,\,
\,\,\,\,\, \,\,\,\,\, \,\,\,\,\, %\,\,\,\,\, 
\,\,\,\,\, \,\,\,\,\, \,\,\,\,\, %\,\,\,\,\,  
 \times \min \Biggr\{ \sum_{\varW'\in \cndV_{\varN}(Q) :\, H(\varW'|Q) \le H(\varW|Q)} \frac{\cardVsh{Q}{\varW'}}{\dmn^{\varN(\varkin-kR)-1}}, \ 1\ \Biggl\}\\
& \le & \dmn \sum_{Q\in\tcQ_{\varN}} \sum_{\varW \in \cndV_{\varN}(Q)} \exp_{\dmn} [ -\varN
 D(Q\pdsm \varW || \Pin) ] \\
&&\,\,\,\,\, \,\,\,\,\, \,\,\,\,\, \,\,\,\,\, 
\,\,\,\,\, \,\,\,\,\, \,\,\,\,\, %\,\,\,\,\, 
\times \sum_{\varW'\in \cndV_{\varN}(Q) :\, H(\varW'|Q)\le H(\varW|Q)} \exp_{\dmn} [ -\varN
 |\varkin-kR-H(\varW'|Q)|^{+} ]\\
& \le & \dmn \sum_{Q\in\tcQ_{\varN}} \sum_{\varW \in \cndV_{\varN}(Q)} 
\exp_{\dmn} [ -\varN  D(Q\pdsm \varW || \Pin) ] \\ %\, %\times 
&&\,\,\,\,\, \,\,\,\,\, \,\,\,\,\, \,\,\,\,\, 
\,\,\,\,\, \,\,\,\,\, \,\,\,\,\, %\,\,\,\,\, 
\times |\cndV_{\varN}(Q)| \max_{\varW'\in\cndV_{\varN}(Q) :\, H(\varW'|Q) \le H(\varW|Q)
}\exp_{\dmn} [ -\varN |\varkin-kR-H(\varW'|Q)|^{+} ]\\
& = &\dmn \sum_{Q\in\tcQ_{\varN}} \sum_{\varW \in \cndV_{\varN}(Q)} 
\exp_{\dmn} [ -\varN  D(Q\pdsm \varW || \Pin) ] \,
|\cndV_{\varN}(Q)| \exp_{\dmn} [ -\varN |\varkin-kR-H(\varW|Q)|^{+} ]\\
& \le & \dmn \sum_{Q\in\tcQ_{\varN}} |\cndV_{\varN}(Q)|^2 \max_{\varW\in\cndV_{\varN}(Q)} \exp_{\dmn} 
[ -\varN D(Q\pdsm \varW||\Pin) -\varN|\varkin-kR-H(\varW|Q)|^+ ]\\
& \le & \dmn |\tcQ_{\varN}| \cardcndV^2 \max_{Q\in \tcQ_{\varN},\,\varW\in\cndV_{\varN}(Q)} \exp_{\dmn} 
[ -\varN D(Q\pdsm \varW||\Pin) -\varN|\varkin-kR-H(\varW|Q)|^+ ]\\
&\le & \dmn |\tcQ_{\varN}| \cardcndV^2 \exp_{\dmn} \big\{ -\varN \min_{P'\in\tcQ_{\varN}(\myF^{\varnin-\varkin}\times \myF^{2\varkin})} 
[D(P'||\Pin) + |\varkin-kR-H(\overleftarrow{P'}|\Bar{P'})|^+ ] \big\} \\
&= & \dmn |\tcQ_{\varN}| \cardcndV^2 \exp_{\dmn} \big\{ -\varN \min_{P'}
[D(P'||\Pin) + |\varkin-kR-H(\overleftarrow{P'}|\Bar{P'})|^+ ] \big\}\\
&= & \dmn |\tcQ_{\varN}| \cardcndV^2 \exp_{\dmn}[-\varN \Enk{\varnin}{\varkin}(R,\Pin)].
\end{eqnarray*}
This bound on $1-\Fbar$ is trivially true for $R\ge 1$. %$R\ge k$.
Note that $|\tcQ_{\varN}|$ and $\cardcndV$ are polynomial in $\varN$.
We see the bound in the theorem
upon putting $R=R'\varnin/k$. % and $m=\varnin\varN$.

\mysectionapp{Proof of Lemma~\ref{lem:Pin_Ic}  \label{ss:proof4Lemma5}}

Let $\rho=(\dim \Hcd)^{-1}\Pi_{\Hcd}$ and assume $\Hcd=\Hcd^{(0)}$
without loss of generality as in Section~\ref{subsec:te}.
To prove the lemma, we will show two equalities
\begin{equation}\label{eq:S1}
\vNe\big(\cA^{\tnsr n}(\rho)\big)=H(\Bar{\Pin})+\varkin
\end{equation}
and 
\begin{equation}\label{eq:S2}
\vNe\big([\Id \tnsr \cA^{\tnsr n}](\ket{\Psi}\bra{\Psi})\big)=
H(\Bar{\Pin}) + H(\overleftarrow{\Pin}|\Bar{\Pin}),
\end{equation}
where $\ket{\Psi}$ is a purification of $\rho$,
which will establish the statement.

The interpretation of errors $\Ebe_{x}$, $x\in\myF^{2n}$,
in terms of the basis $\{ g_1,h_1,\dots,g_{n},h_{n} \}$
in Section~\ref{subsec:te} is useful to see %the equalities
(\ref{eq:S1}) and (\ref{eq:S2}).
Namely, we trace the action of an error $\Ebe_{x}$,
which can be viewed as $\Bar{X}_{u} \Bar{Z}_{u'} \Ebeh{\vartpr} \Ebeg{t}$
as discussed in Section~\ref{subsec:te}.
%The equality $\vNe\big(\cA^{\tnsr n}(\rho)\big)=H(\Bar{\Pin})+\varkin$
Equation (\ref{eq:S1})
holds because $(\dim \Hcd)^{-1} \Pi_{\Hcd}$ is conveyed
to $(\dim \Hcd^{(s)})^{-1} \Pi_{\Hcd^{(s)}}$ by $\Ebeh{\vartpr}$
with probability $\Bar{\Pin}(s)$,
where $\coch(\varrc)_{2i}=s_i$ for $1\le i \le n-k$
and $\Hcd^{(s)}$ has been given in (\ref{eq:codespace2}) or (\ref{eq:synd2}),
the subspaces $\Hcd^{(s)}$, $s\in\myF^{n-k}$, are mutually orthogonal,
and the action of $\Bar{X}_{u} \Bar{Z}_{u'}$
is similar to that of a tensor product of Pauli matrices or Weyl unitaries,
which leaves the operator $\Pi_{\Hcd^{(s)}}$ unchanged. 

Similar reasoning results in (\ref{eq:S2}).
%$\vNe\big([\Id \tnsr \cA^{\tnsr n}](\ket{\Psi}\bra{\Psi})\big)=H(\Pin)$.
In this case, we trace the action of errors $I\tnsr\Ebe_{x}$
on the state $\ket{\Psi}\bra{\Psi}$, where
\[
\ket{\Psi} = \frac{1}{\dmn^{k/2}} \sum_{(\varb_1,\dots,\varb_k)\in\myF^{k} }
\ket{\ghb{\varb_{1},\dots,\varb_k}} 
\tnsr \ket{\ghb{0,\dots,0,\varb_{1},\dots,\varb_k}}
\]
is a purification of $(\dim \Hcd)^{-1} \Pi_{\Hcd}$.
The action of $\Ebeh{\vartpr} \Ebeg{t}$
(in fact, $I\tnsr\Ebeh{\vartpr} \Ebeg{t}$ in this case) 
is similar to the previous case.
To see how $\Bar{X}_{u} \Bar{Z}_{u'}$ acts on the states,
we use the next fundamental lemma on CP linear maps.
\begin{lemma}\cite{choi75}.\label{lem:choi}
Let $\Hch'$ be a Hilbert space with an orthonormal basis
$\{ \ket{\baseChoi{0}},\dots,\ket{\baseChoi{{K-1}}} \}$.
A linear map $\cM: \Bop(\Hch') \to \Bop(\Hch')$
is completely positive if and only if
$[\Id \tnsr \cM]( \ket{\Phi} \bra{\Phi} )$
is positive, 
where $\Id$ is the identity map on $\Bop(\Hch')$, and
\[
\ket{\Phi} = \frac{1}{\sqrt{K}} \sum_{0 \le i < K} \ket{\baseChoi{i}} \tnsr \ket{\baseChoi{i}}.
\]
Moreover,
if we represent $[\Id \tnsr \cM]( \ket{\Phi} \bra{\Phi} )$ as
\begin{equation*} %\label{eq:chois_matrix2}
[\Id \tnsr \cM]( \ket{\Phi} \bra{\Phi} ) 
\mateq \frac{1}{K} \sum_{x\in\cY} \mbm{m}_{x} \mbm{m}_{x}^{\dagger},
\end{equation*}
where $\mateq$ indicates that the right-hand side is the matrix
of the operator on the left-hand side with respect to the basis 
$\{ \ket{\baseChoi{i}} \tnsr \ket{\baseChoi{j}} \}_{(i,j)\in \{0,\dots,K-1\}^2 }$,
i.e., 
\[
T \mateq (t_{ij,kl})_{(i,j,k,l)\in \{0,\dots,K-1\}^4}\,
\longleftrightarrow\,
T=\sum_{(i,j,k,l)\in \{0,\dots,K-1\}^4} t_{ij,kl}
(\ket{\baseChoi{i}} \tnsr \ket{\baseChoi{j}})
(\bra{\baseChoi{k}} \tnsr \bra{\baseChoi{l}}),
\]
and rearrange the elements of 
\[
\mbm{m}_{x}=(\varm_{x,0,0},\dots,\varm_{x,0,K-1},\dots,
\varm_{x,K-1,0},\dots,\varm_{x,K-1,K-1})\transp \in 
\bC^{K^2}
\]
into the matrix form 
\[
 \hat{M}_{x}= \begin{pmatrix}\varm_{x,0,0} & \dots & \varm_{x,K-1,0}\\
\vdots\ & & \vdots\ \\
\varm_{x,0,K-1} & \dots &\varm_{x,K-1,K-1}
\end{pmatrix},\quad x\in\cY,
\]
then we obtain an operator-sum representation of $\cM$:
$\cM \sim \{ M_{x} \}_{x\in\cY}$,
where $\hat{M}_x$ is the matrix of $M_{x}$ with respect to the basis 
$\{ \ket{\baseChoi{i}} \}$, $x\in\cY$.
\enlem

Due to Lemma~\ref{lem:te}, the matrix
of $[\Id \tnsr \cA^{\tnsr n}](\ket{\Psi}\bra{\Psi})$
with respect to the basis that consists of
\begin{gather*}
\ket{\ghb{\varb_{1},\dots,\varb_k}} 
\tnsr \ket{\ghb{s_1,\dots,s_{n-k},\varb'_{1},\dots,\varb'_k}},\\
(s_1,\dots,s_{n-k})\in\myF^{n-k},\ (\varb_{1},\dots,\varb_k),
(\varb'_{1},\dots,\varb'_{k}) \in\myF^{k},
\end{gather*}
is block diagonal [where the basis elements are arranged in a
lexicographic order on $(s_1,\dots,s_{n-k};$ %$LAYOUT
$\varb_1,\dots,\varb_{k};\varb'_1,\dots,\varb'_{k})$], 
and owing to Lemma~\ref{lem:choi},
which we apply putting $K=\dmn^{k}$ and 
\[ 
\ket{\Phi}=
\dmn^{-k/2} \sum_{(b_1,\dots,b_k)\in\myF^{k}}
\ket{\ghb{\varb_{1},\dots,\varb_k}} 
\tnsr \ket{\ghb{s_1,\dots,s_{n-k},\varb_{1},\dots,\varb_k}}
\]
to each block,
von Neumann entropy of the block with label 
$s=(s_1,\dots,s_{n-k})$, after normalization,
equals Shannon entropy of $\rvY$ conditional on $\rvX=s$,
where the pair of random variables $(\rvX,\rvY)$ is drawn according to $\Pin$.
Thus, we have (\ref{eq:S2}), completing the proof.

\mysectionapp{Proof of Lemma~\ref{lem:ub4cond_cap} \label{ss:proof4ub}}

%{\em Proof.}\/
Barnum {\em et al.}~\cite[p.~4162]{barnum98} have shown the inequality
\begin{equation}\label{eq:proof1}
\vNe(\rho) \le I_{\rm c}(\rho, \cA^{\tnsr n}) +2 + 4 [ 1- F_{\rm e}( \rho, \cR_n \cA^{\tnsr n})] n,
\end{equation}
which holds for any state $\rho$ in $\Bop(\Hch^{\tnsr n})$, channel $\cA$ and
TPCP linear map $\cR_n:\, \Bop(\Hch^{\tnsr n}) \to \Bop(\Hch^{\tnsr n})$,
where $F_{\rm e}$ denotes the entanglement fidelity.
Also it is known that $F(\Hcd,\cR_n\cA^{\tnsr n}) \ge 1-\eta$
implies $F_{\rm e}\big((\dim \Hcd)^{-1} \Pi_{\Hcd},\cR_n\cA^{\tnsr n}\big) \ge 1-(3/2)\eta$~\cite[p.~1324, Theorem~2]{barnum00}. 
Putting $\rho=(\dim \Hcd_n)^{-1} \Pi_{\Hcd_n}$ in 
(\ref{eq:proof1}) and assuming $\Hcd_n\in\rsCgen_n$ and 
$F(\Hcd_n,\cR_n\cA^{\tnsr n}) \to 1$
as $n$ goes to infinity, we have
\[
\limsup_{n\to\infty} \frac{\log_{\dmn} \dim \Hcd_n}{n} \le \limsup_{n\to\infty}
\sup_{\Hcd\in \rsCgen_n}
\frac{I_{\rm c}\big((\dim \Hcd)^{-1} \Pi_{\Hcd}, \cA^{\tnsr n}\big)}{n},
\] 
and hence, the lemma.

\mysectionapp{Proof of Lemma~\ref{lem:cond_stab} \label{ss:proof_cond}}

First, note that in
the proof of Theorem~\ref{th:main}
(Section~\ref{ss:proof} and Appendix~\ref{ss:proof4Lemma5}), 
we have assumed that the operator basis $\Ebasis=\{ \Ebe_{u}\}_{u\in\cX}$
employed for code design is exactly the same as that
used in the representation
$\{ \sqrt{P(u)}\Ebe_{u} \}_{u\in\cX}$ of the Pauli channel,
and that the proof %of Theorem~\ref{th:main} 
has actually shown
\begin{equation}
\Capa(\cA|\{ \rsC_n(\Ebasis) \}) 
\ge \sup_{\varnin \ge 1}
\max_{ \Hcd \in \rsC_{\varnin}(\Ebasis)}
\frac{I_{\rm c}\big((\dim \Hcd)^{-1} \Pi_{\Hcd},\cA^{\tnsr \varnin}\big)}{\varnin} \label{eq:conditional_lb}
\end{equation}
for %any $\Ebe$ and 
any Pauli or $\Ebasis$-channel $\cA \sim \{ \sqrt{P(u)} \Ebe_{u} \}_{u\in\cX}$.

Put
\[ %\begin{equation}
c_n = \max_{\Hcd \in \rsC_{\varnin}(\Ebasis) }
I_{\rm c}\big((\dim \Hcd)^{-1} \Pi_{\Hcd}, \cA^{\tnsr \varnin}\big)
=\max_{\Cso} [\varkin-H(\overleftarrow{\Pin}|\Bar{\Pin})],
\quad n=1,2,\dots,
\] %\end{equation}
where the second equality is due to Lemma~\ref{lem:Pin_Ic}.
From (\ref{eq:conditional_lb}) and Lemma~\ref{lem:ub4cond_cap} with
$\rsCgen_n=\rsC_{\varnin}(\Ebasis)$, we will obtain the lemma
if we show that the limit of $c_n/n$ exists and 
\begin{equation}\label{eq:superadditive2}
\lim_{n\to\infty}c_n/n =\sup_{n \ge 1}c_n/n.
\end{equation}
To do this, we will show
\begin{equation}\label{eq:superadditive}
c_{n+n'} \ge c_n+c_{n'}.
\end{equation}
The fact that (\ref{eq:superadditive}), together with
the boundedness of $c_n/n$, implies (\ref{eq:superadditive2})
for a general sequence of real numbers $\{ c_n \}$ has often been used 
in (quantum) information theory~\cite{holevo78,barnum98,ogawaPhD}. %ver2

Now let $\Cso\subset\myF^{2n}$ and $\Cso'\subset\myF^{2n'}$ 
with $\dim \Cso=n-k$ and $\dim \Cso'=n'-k'$ achieve the maxima
of $k-H(\overleftarrow{\Pin}|\Bar{\Pin})$
and $k'-H(\overleftarrow{P_{\Cso'}}|\Bar{P_{\Cso'}})$, respectively.
Recall that $H(\overleftarrow{\Pin}|\Bar{\Pin})$ and
$H(\overleftarrow{P_{\Cso'}}|\Bar{P_{\Cso'}})$
are determined from
coset arrays of $\Cso$ and $\Cso'$ defined in Section~\ref{ss:ca}.
All we have to show is the existence of a self-orthogonal
subspace $\Cso'' \subset \myF^{2(n+n')}$ with $\dim \Cso''=n+n'-k''$
such that
\[
k''- H(\overleftarrow{P_{\Cso''}}|\Bar{P_{\Cso''}})
\ge k- H(\overleftarrow{P_{\Cso}}|\Bar{P_{\Cso}})
+k' - H(\overleftarrow{P_{\Cso'}}|\Bar{P_{\Cso'}}).
\]
We can see that
\[
\Cso\Cso' \defeq \{ xy \in\myF^{n+n'} \mid x\in\Cso,y\in\Cso' \}
\]
with $\dim \Cso\Cso'=n+n'-(k+k')$, where $xy$ denotes the
vector obtained by pasting %(concatenating) 
$x$ and $y$ together,
%in this order,
is such a code as follows.
We consider probability arrays of $\Cso$ and $\Cso'$ as in (\ref{eq:carray2}).
Then, it is easy to see that the $\dmn^{n+n'-k-k'} \times \dmn^{2(k+k')}$
array whose $(s,u)$-entry is $P_{\Cso\Cso'}(ss',uu')=\Pin(s,u) \Pinpr(s',u')$, 
$(s,u)\in\myF^{n-k}\times\myF^{2k}$, $(s',u')\in\myF^{n'-k'}\times\myF^{2k'}$,
is a probability array of $\Cso\Cso'$.
From this array, we have
\begin{eqnarray*}
\lefteqn{ k+k'- H(\overleftarrow{P_{\Cso\Cso'}}|\Bar{P_{\Cso\Cso'}})}\\
&=& k+k'- H(P_{\Cso\Cso'}) + H(\Bar{P_{\Cso\Cso'}})\\
&=& k+k'- H(P_{\Cso})- H(P_{\Cso'}) +H(\Bar{P_{\Cso}}) +H(\Bar{P_{\Cso'}})\\
&=& k- H(\overleftarrow{P_{\Cso}}|\Bar{P_{\Cso}})+k'- H(\overleftarrow{P_{\Cso'}}|\Bar{P_{\Cso'}}).
\end{eqnarray*}
[To see these equalities, introduce random variables
$\rvX,\rvY,\rvX',\rvY'$ such that 
$\Prob \{ \rvX=s,\rvY=u,\rvX'=s',\rvY'=u' \} = \Pin(s,u)P_{\Cso'}(s',u')$.]
Hence, we have (\ref{eq:superadditive}) and consequently the lemma. 
%proposition.


\begin{thebibliography}{10}

\bibitem{shor95}
P.~W. Shor, ``Scheme for reducing decoherence in quantum computer memory,''
  {\em Phys.\ Rev.\ A}, vol.~52, pp.~R2493--2496, 1995.

\bibitem{barnum98}
H.~Barnum, M.~A. Nielsen, and B.~Schumacher, ``Information transmission through
  a noisy quantum channel,'' {\em Phys.\ Rev.\ A}, vol.~57, pp.~4153--4175,
  June 1998.

\bibitem{barnum00}
H.~Barnum, E.~Knill, and M.~A. Nielsen, ``On quantum fidelities and channel
  capacities,'' {\em IEEE Trans. Information Theory}, vol.~46, pp.~1317--1329,
  July 2000.

\bibitem{HolevoWerner01}
A.~S. Holevo and R.~F. Werner, ``Evaluating capacities of bosonic {G}ausssian
  channels,'' {\em Phys.\ Rev.\ A}, vol.~63, no.~3, pp.~032312--1--14, 2001.

\bibitem{holevo01s}
A.~S. Holevo, {\em Statistical Structure of Quantum Theory}.
\newblock Berlin: Springer, 2001.

\bibitem{HPreskill01}
J.~Harrington and J.~Preskill, ``Achievable rates for the {G}aussian quantum
  channel,'' {\em Phys.\ Rev.\ A}, vol.~64, pp.~062301--1--9, 2001.

\bibitem{bennett96m}
C.~H. Bennett, D.~P. DiVincenzo, J.~A. Smolin, and W.~K. Wootters,
  ``Mixed-state entanglement and quantum error correction,'' {\em Phys.\ Rev.\
  A}, vol.~54, pp.~3824--3851, Nov. 1996.

\bibitem{gottesmanPhD}
D.~Gottesman, {\em Stabilizer Codes and Quantum Error Correction}.
\newblock {Ph.D.} thesis, California Institute of Technology, May 1997.
\newblock E-print, quant-ph/9705052, LANL, 1997.

\bibitem{preskillLNbook0}
J.~Preskill, {\em Lecture Notes for Physics 229: Quantum Information and
  Computation}.
\newblock \texttt{http://www.theory.caltech.edu/people/preskill/ph229}, 1998.

\bibitem{hamada01e}
M.~Hamada, ``Exponential lower bound on the highest fidelity achievable by
  quantum error-correcting codes,'' {\em Phys.\ Rev.\ A}, vol.~65,
  pp.~052305--1--4, 2002.
\newblock E-Print, quant-ph/0109114, LANL, 2001.

\bibitem{hamada01g}
M.~Hamada, ``Lower bounds on the quantum capacity and highest error exponent of
  general memoryless channels,'' {\em IEEE Trans.\ Information Theory},
  vol.~48, no.~9, pp.~2547--2557, 2002.
\newblock E-Print, quant-ph/0112103, LANL, 2001.


\bibitem{ss97}
P.~W. Shor and J.~A. Smolin, ``Quantum error-correcting codes need not
  completely reveal the error syndrome,'' e-Print quant-ph/9604006, LANL, 1996.

\bibitem{dss98}
D.~P. DiVincenzo, P.~W. Shor, and J.~A. Smolin, ``Quantum-channel capacity of
  very noisy channels,'' {\em Phys.\ Rev.\ A}, vol.~57, pp.~830--839, Feb.
  1998.
\newblock Correction: Phys.\ Rev.\ A, 59, p.~1717.

\bibitem{csiszar_koerner}
I.~Csisz\'{a}r and J.~K\"{o}rner, {\em Information Theory: Coding Theorems for
  Discrete Memoryless Systems}.
\newblock NY: Academic, 1981.

\bibitem{csiszar98}
I.~Csisz\'{a}r, ``The method of types,'' {\em IEEE Trans. Information Theory},
  vol.~IT-44, pp.~2505--2523, Oct. 1998.

\bibitem{shor95f}
P.~W. Shor, ``Fault-tolerant quantum computation,'' in {\em 37th Symposium on
  Foundations of Computer Science}, pp.~56--65, IEEE Computer Society Press,
  1996.
\newblock qaunt-ph/9605011.

\bibitem{steane99}
A.~M. Steane, ``Efficient fault-tolerant quantum computing,'' {\em Nature},
  vol.~399, pp.~124--126, 1999.

\bibitem{forney}
J.~G.~D.~Forney, {\em Concatenated Codes}.
\newblock MA: MIT Press, 1966.

\bibitem{gottesman96}
D.~Gottesman, ``Class of quantum error-correcting codes saturating the quantum
  {H}amming bound,'' {\em Phys.\ Rev.\ A}, vol.~54, pp.~1862--1868, Sept. 1996.

\bibitem{crss97}
A.~R. Calderbank, E.~M. Rains, P.~W. Shor, and N.~J.~A. Sloane, ``Quantum error
  correction and orthogonal geometry,'' {\em Phys.\ Rev.\ Lett.}, vol.~78,
  pp.~405--408, Jan. 1997.

\bibitem{crss98}
A.~R. Calderbank, E.~M. Rains, P.~W. Shor, and N.~J.~A. Sloane, ``Quantum error
  correction via codes over {GF(4)},'' {\em IEEE Trans.\ Inform.\ Theory},
  vol.~44, pp.~1369--1387, July 1998.

\bibitem{rains99}
E.~M. Rains, ``Nonbinary quantum codes,'' {\em IEEE Trans.\ Information
  Theory}, vol.~45, pp.~1827--1832, Sept. 1999.

\bibitem{GottesmanKP01}
D.~Gottesman, A.~Kitaev, and J.~Preskill, ``Encoding a qubit in an
  oscillator,'' {\em Phys.\ Rev.\ A}, vol.~64, pp.~012310--1--21, 2001.

\bibitem{ABargKnillL00}
A.~E. Ashikhmin, A.~M. Barg, E.~Knill, and S.~N. Litsyn, ``Quantum error
  detection {I} and {II},'' {\em IEEE Trans. Information Theory}, vol.~46,
  pp.~778--800, May 2000.

\bibitem{FNagaoka98}
A.~Fujiwara and H.~Nagaoka, ``Operational capacity and pseudoclassicality of a
  quantum channels,'' {\em IEEE Trans.\ Inform.\ Theory}, vol.~44,
  pp.~1071--1086, May 1998.

\bibitem{king01a}
C.~King, ``Additivity for a class of unital qubit channels,'' e-Print
  qunat-ph/0103156, LANL, 2001.

\bibitem{king02}
C.~King, ``The capacity of the quantum depolarizing channel,'' e-Print
  qunat-ph/0204172, LANL, 2002.

\bibitem{AHolevo01}
G.~G. Amosov and A.~S. Holevo, ``On the multiplicativity conjecture for quantum
  channels,'' e-Print qunat-ph/0103015, LANL, 2001.

\bibitem{shor02a}
P.~W. Shor, ``Additivity of the classical capacity of entanglement-breaking
  quantum channels,'' e-Print qunat-ph/0201149, LANL, 2002.

\bibitem{bs01ea}
C.~H. Bennett, P.~W. Shor, J.~A. Smolin, and A.~V. Thapliyal,
  ``Entanglement-assisted capacity of a quantum channel and the reverse {S}hannon
  theorem,'' e-Print qunat-ph/0106052, LANL, 2001.

\bibitem{holevo01ea}
A.~S. Holevo, ``On entanglement-assisted classical capacity,'' e-Print
  qunat-ph/0106075, LANL, 2001.

\bibitem{KeylWerner02}
M.~Keyl and R.~F. Werner, ``How to correct small quantum errors,'' e-Print
  quant-ph/0206086, LANL, 2002.

\bibitem{artin}
E.~Artin, {\em Geometric Algebra}.
\newblock New York: Interscience Publisher, 1957.

\bibitem{grove}
L.~C. Grove, {\em Classical Groups and Geometric Algebra}.
\newblock Providence, Rhode Island: American Mathematical Society, 2001.

\bibitem{kraus71}
K.~Kraus, ``General state changes in quantum theory,'' {\em Annals of Physics},
  vol.~64, pp.~311--335, 1971.

\bibitem{choi75}
M.-D. Choi, ``Completely positive linear maps on complex matrices,'' {\em
  Linear Algebra and Its Applications}, vol.~10, pp.~285--290, 1975.

\bibitem{schumacher96}
B.~Schumacher, ``Sending entanglement through noisy quantum channels,'' {\em
  Phys.\ Rev.\ A}, vol.~54, pp.~2614--2628, Oct. 1996.

\bibitem{nielsen_chuang}
M.~A. Nielsen and I.~L. Chuang, {\em Quantum Computation and Quantum
  Information}.
\newblock Cambridge, UK: Cambridge University Press, 2000.

\bibitem{KnillLaflamme97}
E.~Knill and R.~Laflamme, ``Theory of quantum error-correcting codes,'' {\em
  Phys.\ Rev.\ A}, vol.~55, pp.~900--911, Feb. 1997.

\bibitem{weyl31}
H.~Weyl, {\em The Theory of Groups and Quantum Mechanics}.
\newblock NY: Dover, 1950.
\newblock Translation from the second German ed., 1931.

\bibitem{schwinger60}
J.~Schwinger, ``Unitary operator bases,'' {\em Proc.\ Nat.\ Acad.\ Sci.\ USA},
  vol.~46, pp.~570--579, 1960.

\bibitem{knill96a}
E.~Knill, ``Non-binary unitary error bases and quantum codes,'' e-Print
  quant-ph/9608048, LANL, 1996.

\bibitem{knill96b}
E.~Knill, ``Group representations, error bases and quantum codes,'' e-Print
  quant-ph/9608049, LANL, 1996.

\bibitem{BennettShor98}
C.~H. Bennett and P.~W. Shor, ``Quantum information theory,'' {\em IEEE Trans.
  Information Theory}, vol.~44, pp.~2724--2742, Oct. 1998.

\bibitem{AshikhminKnill00}
A.~Ashikhmin and E.~Knill, ``Nonbinary quantum stabilizer codes,'' {\em IEEE
  Trans. Information Theory}, vol.~47, pp.~3065--3072, Nov. 2001.

\bibitem{gottesman99}
D.~Gottesman, ``Fault-tolerant quantum computation with higher-dimensional
  systems,'' {\em Lecture Notes in Comp.\ Sci.}, vol.~1509, pp.~302--313, 1999.

\bibitem{takeuchi}
G.~Takeuchi, {\em Senkei-Daisu to Ryoshi-Rikigaku (Linear Algebra and Quantum
  Mechanics)}.
\newblock Tokyo: Shokabo, 1981.

\bibitem{slepian56}
D.~Slepian, ``A class of binary signaling alphabets,'' {\em The Bell System
  Technical Journal}, vol.~35, pp.~203--234, Jan. 1956.
\newblock Reprinted in E.\ R.\ Berlekamp, ed., {\em Key Papers in The
  Development of Coding Theory},\/ NY, IEEE Press, 1974.

\bibitem{ptrsn}
W.~W. Peterson and E.~J. {Weldon, Jr.}, {\em Error-Correcting Codes}.
\newblock MA: MIT Press, 2nd~ed., 1972.

\bibitem{gabidulin67}
E.~M. Gabidulin, ``Limits for the decoding error probability when linear codes
  are used in memoryless channels,'' {\em Problems of Information
  Transmission}, vol.~3, no.~2, pp.~43--48, 1967.

\bibitem{werner01}
R.~F. Werner, ``All teleportation and dense coding scheme,'' {\em J.\ Phys.\ A:
  Math.\ Gen.}, vol.~34, pp.~7081--7094, 2001.

\bibitem{barnum98e}
H.~Barnum, J.~A. Smolin, and B.~M. Terhal, ``Quantum capacity is properly
  defined without encodings,'' {\em Phys.\ Rev.\ A}, vol.~58, pp.~3496--3501,
  Nov. 1998.

\bibitem{barg02}
A.~Barg, ``A low-rate bound on the reliability of a quantum discrete memoryless
  channel,'' e-Print quant-ph/0203077, LANL, 2002.

\bibitem{holevo78}
A.~S. Holevo, ``Capacity of a quantum communications channel,'' {\em Problems
  of Information Transmission}, vol.~15, pp.~247--253, Oct.--Dec. 1979.

\bibitem{ogawaPhD}
T.~Ogawa, {\em A study on the asymptotic property of the hypothesis testing and
  the channel coding in quantum mechanical systems}.
\newblock {Ph.D.} thesis, University of Electro-Communications, Chofu-shi,
  Tokyo, 2000.
\newblock In Japanese.

\end{thebibliography}
\end{document}